\documentclass[twocolumn,english,aps,unsortedaddress,pre,showpacs]{revtex4}
\usepackage[T2A,T1]{fontenc}
\usepackage[latin9]{inputenc}
\usepackage{bm}
\usepackage{amsmath}
\usepackage{amssymb}
\usepackage{graphicx}
\usepackage{esint}

\makeatletter

\providecommand{\tabularnewline}{\\}

\@ifundefined{textcolor}{}
{%
 \definecolor{BLACK}{gray}{0}
 \definecolor{WHITE}{gray}{1}
 \definecolor{RED}{rgb}{1,0,0}
 \definecolor{GREEN}{rgb}{0,1,0}
 \definecolor{BLUE}{rgb}{0,0,1}
 \definecolor{CYAN}{cmyk}{1,0,0,0}
 \definecolor{MAGENTA}{cmyk}{0,1,0,0}
 \definecolor{YELLOW}{cmyk}{0,0,1,0}
}

\usepackage{bm}

\usepackage{babel}

\makeatother

\usepackage{babel}
\begin{document}

\preprint{This line only printed with preprint option}

\title{Emergence of jams in the generalized totally asymmetric simple exclusion
process}

\author{A.E. Derbyshev}

\email{derbishev@theor.jinr.ru}

\selectlanguage{english}%

\affiliation{Bogoliubov Laboratory of Theoretical Physics, Joint Institute for
Nuclear Research, Dubna, Russia}

\affiliation{Moscow Institute of Physics and Technology, Dolgoprudny, Russia }

\author{A.M.Povolotsky}

\email{alexander.povolotsky@gmail.com}

\selectlanguage{english}%

\affiliation{Bogoliubov Laboratory of Theoretical Physics, Joint Institute for
Nuclear Research, 141980 Dubna, Russia}

\affiliation{Laboratory of Mathematical Physics, NRU HSE, Moscow, Russia }

\author{V. B. Priezzhev}

\email{priezzvb@theor.jinr.ru}

\selectlanguage{english}%

\affiliation{Bogoliubov Laboratory of Theoretical Physics, Joint Institute for
Nuclear Research, 141980 Dubna, Russia}

\begin{abstract}
The generalized totally asymmetric exclusion process (TASEP) {[}J.
Stat. Mech. P05014 (2012){]} is an integrable generalization of the
TASEP equipped with an interaction, which enhances the clustering
of particles. The process interpolates between two extremal cases:
the TASEP with parallel update and the process with all particles
irreversibly merging into a single cluster moving as an isolated particle.
We are interested in the large time behavior of this process on a
ring in the whole range of the parameter $\lambda$ controlling the
interaction. We study the stationary state correlations, the cluster
size distribution and the large-time fluctuations of integrated particle
current. When $\lambda$ is finite, we find the usual TASEP-like behavior:
The correlation length is finite; there are only clusters of finite
size in the stationary state and current fluctuations belong to the
Kardar-Parisi-Zhang universality class. When $\lambda$ grows with
the system size so does the correlation length. We find a nontrivial
transition regime with clusters of all sizes on the lattice. We identify
a crossover parameter and derive the large deviation function for
particle current, which interpolates between the case considered by
Derrida-Lebowitz and a single particle diffusion. 
\end{abstract}
\maketitle

\section{Introduction\label{sec:Introduction}}

The asymmetric simple exclusion process (ASEP) is one of the basic
models of driven transport admitting an analytical treatment \cite{Spoh91,Ligg99,Gunter}.
It is commonly accepted that different versions of ASEP provide an
adequate description of statistical properties of one-dimensional
diffusive and driven-diffusive systems. During the last decades the
ASEP was a laboratory for obtaining the universal critical exponents
and scaling functions of the Edward-Wilkinson (EW) and Kardar-Parisi-Zhang
(KPZ) universality classes \cite{EW,KPZ}. The range of models that
can be solved exactly is very limited, but the universality implies
that results obtained from their solution apply to a wide range of
stochastic systems, like interacting particle systems, growing interfaces,
crystal facets, polymers in random media, etc. Among the results,
which are believed to be universal, are the dynamical exponent of
the KPZ class \cite{Gwa Spohn}, the KPZ-EW crossover function for
the relaxation time \cite{Doochul Kim}, the large deviation function
(LDF) for particle current in the systems with periodic \cite{Derrida Lebowitz}
and open \cite{de Gier Essler,GLMV} boundary conditions. More recently,
consideration of these processes on the infinite lattice yielded plenty
of results on the universal scaling functions for probability distributions
and correlation functions characterizing the nonstationary time evolution.

The totally asymmetric exclusion process (TASEP) is the simplest version
of the ASEP, possessing a special mathematical structure, which simplifies
the solution significantly. Using this structure, Derrida and Lebowitz
obtained the first exact expression for the LDF of particle current
for an arbitrary lattice size, which yielded the universal scaling
function in the scaling limit. Also, closed determinantal formulas
for the Green's functions were derived for the TASEP on both the infinite
lattice \cite{Schutz} and the ring \cite{priezzhev}. Finally, all
multipoint correlation functions for the process in the infinite system
were constructed \cite{Sasamoto,BFPS,IS,BFPS1,BF,BFS,PovPriS,PPP}.
Remarkably, unlike the partially asymmetric case, the TASEP remains
exactly solvable in a discrete time framework. The models with several
different updates were solved: backward sequential \cite{BrPrS},
parallel \cite{PP} and sublattice parallel \cite{PPS}. All these
versions of the TASEP demonstrate the same universal KPZ behavior
in the scaling limit. It is of interest, however, to examine possible
mechanisms taking the system away from the KPZ class, to see how the
KPZ universality breaks down.

To our knowledge, the generalized TASEP (gTASEP) studied here was
first considered in \cite{Woelki}, where without any reference to
its integrable structure, it was used as an example of a traffic model
with a stationary measure admitting factorized representation. It
was later rediscovered in \cite{genTASEP}, within a totally different
context as an integrable generalization of the TASEP. Finally, it
was shown to be a particular $q=0$ limiting case of the general three
parametric Bethe ansatz-solvable stochastic chipping model \cite{chipping},
also referred to as a q-Hahn or $(q,\mu\text{,\ensuremath{\nu}})-$boson
process \cite{Corwin2,BCPS}. In turn, it containes already known
TASEPs with parallel and sequential update as particular cases. In
the gTASEP an additional interaction between particles is introduced,
which enhances the clustering of particles comparing to the usual
TASEP. The dynamics of the model can be viewed as the TASEP-like process,
where clusters of particles diffuse, breaking into parts and merging
together. The relative frequency of these processes is controlled
by an extra parameter $\lambda$. The bigger value of $\lambda$,
the stronger is the effective attraction between particles and the
larger is the size of clusters in typical particle configuration.
A limiting case $\lambda\to\infty,$ which we refer to as the deterministic
aggregation (DA) limit, produces the process, where particles stick
together irreversibly, finally forming a single giant cluster, which
moves as an ordinary random walk.

The main aim of the present paper is to study how the large-scale
behavior of the steady state in gTASEP changes as the DA limit is
approached. We concentrate on the stationary state correlations and
fluctuations of particle current on the ring. For moderate interaction
strength it is natural to expect that the scaling behavior of gTASEP
will be similar to the usual TASEP, which belongs to the KPZ universality
class. For the latter, it is well known that the stationary state
is uncorrelated if looked at in the scale of the system size. Also
the motion of particles in an infinite system is subdiffusive. Though
it is still diffusive in a finite system, the diffusion coefficient
decays as $\Delta\sim1/\sqrt{L}$ as the system size $L$ grows to
infinity. Further details of the large time fluctuations of particle
current can be extracted from the universal LDF obtained by Derrida
and Lebowitz in \cite{Derrida Lebowitz} for the usual TASEP, and
later proved to hold for several other systems \cite{lee kim,PPH,povolotsky,Povolotsky Mendes}.
On the other hand in the DA limit the particles form a single giant
cluster, which moves as a single particle. This behavior obviously
corresponds to correlation length unboundedly growing with the system
size and to purely diffusive motion of each individual particle. As
a result, there are many small particle clusters, finite range correlations,
and KPZ-like fluctuations on the one end of the range of $\mbox{\ensuremath{\lambda}}$
and one macroscopic cluster with pure single-particle diffusion on
the other. Then the natural question to ask is how many particle clusters
can there be and how large can typical particle clusters be for the
KPZ universality to survive and how the two regimes are connected
to each other. Intuition says that at least at a finite density of
finite clusters, which is maintained at finite values of $\lambda,$
we should be in the KPZ regime, as the finite clusters can be effectively
treated as larger particles. The analysis below shows, however, that
one can approach the DA limit much more closely keeping the universal
KPZ form of the current LDF. We show that even when $\lambda$ and,
hence, the typical size of clusters, grow with the system size, the
LDF preserves its functional form, unless the order of $\lambda$
is as large as $L^{2}$. When $\lambda/L^{2}\to0$, the dependence
on the value of $\lambda$ affects only the non-universal constants
controlling the fluctuation scale but not the functional form of the
distribution. At the scale $\lambda\sim L^{2}$ there is only a few
(a finite number of) macroscopic clusters on the lattice and the correlation
length is of order of system size $L$. At this scale, the transition
from the KPZ to the DA limit takes place. We obtain the LDF that crosses
over from the KPZ Derrida-Lebowitz form to pure Gaussian as $\lambda/L^{2}$
varies from zero to infinity.

To have a rough idea of where the scale $\lambda\sim L^{2}$ comes
from, the following simple mean-field argument can be used, which
should not be considered as a derivation, but can be viewed as a description
of the scenario of the transtion regime. Let us think about the large
number $M$ of interacting particles diffusing on the one-dimensional
lattice with overall particle density being fixed, $M/L=c$, as about
diffusing, aggregating and dissociating clusters (compact groups of
particles). Two clusters merge when coming in contact, while any cluster
can break down into two smaller clusters at any point with small rate
$\alpha$. Then as $\alpha$ goes to zero, we expect to observe a
transition from a finite density of finite clusters to a single cluster
of size $M$. In the transition regime there is a finite number of
clusters of any macroscopic size. Under this suggestion consider the
conditions for the equilibrium to hold between merging and breaking
up clusters at all scales. Let $P(n)$ be a global density (mean number
per unit length) of $n$-particle clusters, which is supposed to be
of the same order through the whole range of $n$ in the transition
regime. The number of clusters of size $n$ in the system is equal
to $LP(n)$, and the total number of clusters of any size in a typical
configuration is given by the sum $L\sum_{1\leq n\leq M}P(n)$. For
this number to be finite (of order of one), the value of $P(n)$ should
be of order of $1/L^{2}$. As the cluster of $n$ particles can split
into two smaller clusters at any of its points, the mean rate of decay
of such clusters will be $n\alpha P(n)$. On the other hand, the number
of clusters of size $n$ appearing per unit time is $\sum_{k}P(k)P(n-k)$,
which is of order of $1/L^{3}.$ Equating these two expressions we
find that $\alpha$ must be of order of $1/L^{2}.$ An analog of splitting
rate $\alpha$ in our model is the inverse of the parameter $\lambda$.
Our asymptotic analysis indeed shows that the scaling parameter controlling
the transition occurring in the limit $\lambda\to\infty$ can be chosen
proportional to $\lambda/L^{2}.$ Tuning this parameter one can obtain
both the particle current LDFs for KPZ and DA regimes as limiting
cases. We want to emphasize that the above description gives a qualitative
picture, which can only illustrate the exact results obtained below.

It is worth mentioning other studies of models, where the particle
clustering strength can be controlled. A version of the TASEP with
next-nearest neighbor interaction was proposed in \cite{AntalSchutz}.
The particle flow has the jamming tendency and for this reason the
flow diagram is shifted in the region of large densities. The finding
of a ``fourth phase'' in the mean-field approximation (approved by
Monte Carlo simulations) demonstrates an unusual and nontrivial character
of particle flow when it enters the jam regime.

The model which allows for the diffusion of clusters, aggregation
on contacts between them and single-particle dissociation has been
considered in \cite{Majumdar}. A mean field analysis of the model
showed that the system undergoes the dynamical phase transition: The
steady state mass distribution in one phase decays exponentially for
large masses. In another phase, the model predicts an infinite aggregate
in addition to a power-law mass decay.

Note that the mentioned models do not belong to the class of integrable
models. The models like these are generally studied in the mean field
approximation, or at best allow the exact characterization of the
stationary state distribution; see, e.g., \cite{AntalSchutz}. Such
an analysis provides the thermodynamical description, like the density-current
relation, which is not universal and to large extent depends on particular
dynamical rules. In contrast, in our case the integrability allows
the exact treatment of the full dynamical problem, which contains
information about universal fluctuations in the scaling limit.

Our paper is organized as follows. In Sec. \ref{sec:Model-definition-and}
we formulate the model and explain the zero-ramge rocess (ZRP)-ASEP
mapping, which allows us to establish a relation between gTASEP and
another zero-range type model with an unbounded number of particles
in a site. While many quantities characterizing the two models coincide,
the advantage of models like the ZRP is a factorized form of steady
state distribution, which can be analyzed with the canonical partition
function formalism.

In Sec. \ref{sec:Stationary-state} we study the stationary state
of both gTASEP and the corresponding ZRP-like model. For the ZRP-like
model we obtain the exact expressions for the partition function on
an arbitrary finite lattice and use it to derive the occupation number
distribution. The latter can be reinterpreted as the cluster size
distribution in the gTASEP. We also derive the generating function
of particle jumps and, in particular, obtain the exact formula for
the mean number of particle jumps per unit time. The exact partition
function and particle current are represented as contour integrals,
which, then, are explicitly evaluated in terms of the Gauss $_{2}F_{1}$
and Appell $F_{1}$ terminating hypergeometric series, respectively.
Then we perform an asymptotic analysis of the integrals obtained,
first, in the saddle point approximation, which is applicable when
$\lambda/L^{2}\to0,$ and, second, in the limit $\lambda/L^{2}\to const>0,$
when the saddle point approximation fails. In the first case, we obtain
the geometric finite (or subextensive) cluster size distribution and
the thermodynamic formula for particle current (flow diagram) depending
on two parameters and particle density. In the second case we obtain
the distribution of cluster sizes on the system size scale, expressed
in terms of the modified Bessel functions. In the last section of
Sec. \ref{sec:Stationary-state} we analyze the stationary state of
the gTASEP directly in the grand-canonical ensemble exploiting the
fact that the stationary measure of the gTASEP is similar to the Gibbs
measure of a one-dimensional Ising model. We evaluate the two-point
correlation function and discuss its behavior in both limits.

Section \ref{sec:Statistics-of-particle} is devoted to the analysis
of particle current fluctuations. We first remind the reader of the
Bethe ansatz solution of the models discussed and then obtain the
largest eigenvalue of the Markov matrix deformed by including parameter
$\gamma$, counting the particle jumps. The eigenvalue, obtained in
the parametric form as two series with coefficients expressed via
$_{2}F_{1}$ and $F_{1}$, has a meaning of the rescaled cumulant
generating function of the total number of particle jumps or of the
Legendre transform of the corresponding LDF. In particular, in addition
to the exact particle current obtained in Sec.~\ref{sec:Stationary-state}
we derive the exact expression for the diffusion coefficient of a
particle in gTASEP. The asymptotic analysis again consists of two
parts: the saddle point approximation for $\lambda/L^{2}\to0$, which
reproduces the universal function by Derrida and Lebowitz through
a range of scales of $\lambda$, and the asymptotic analysis on the
scale $\lambda\sim L^{2}$, describing the KPZ-to-Gauss transition.

The last section, Sec. \ref{sec:Universality-and-relation}, is intended
to bind together the variety of the results obtained for the KPZ regime.
We remind to the reader of the scaling theory, developed in \cite{krug meakin halpin-healy},
which claims that many non-universal quantities characterizing the
systems belonging to the KPZ universality class can be expressed in
terms of only two dimensional invariants, which, in particular, are
related to the dimensional constants in the KPZ equation. We show
that this hypothesis is confirmed by our results and conversely express
the non-universal scaling constants in the LDF in terms of the KPZ
dimensional invariants.

In Appendix \ref{sec:Explicit-expressions-of} we give explicit formulas
for some model-dependent constants appearing in the calculations and
establish relations between them, which confirm KPZ universality.
The formulas used to work with the special functions are listed in
Appendices \ref{sec:Hypergeometric-functions} and \ref{sec:Modified-Bessel-function}.

\section{Model definition and ZRP-ASEP mapping\label{sec:Model-definition-and}}

Consider $M$ particles on the one-dimensional lattice of $L$ sites
with periodic boundary conditions. Each lattice site can be occupied
by, at most, one particle. Particle configurations are recorded as
$N$-tuples of particle coordinates $\mathbf{x}=(1\leq x_{1}<x_{2}<\dots<x_{M}\leq L)$.
Particle configurations evolve in discrete time with \textit{clusterwise}
backward-sequential update rules. We refer to a compact string of
particles like $(x_{i-k}=x-k,...,x_{i}=x)$ surrounded by two empty
sites as a cluster. The update of particle configuration at a given
time step starts from the rightmost particle of any cluster. For definiteness,
one can choose the cluster with the maximal coordinate $x_{i}\leq L$
of the rightmost particle. The particle tries to jump one step to
the right, i.e., clockwise, succeeding, $(x\to x+1\mod L$), with
probability $p$ or failing, $(x\to x)$, with probability $1-p$.
In the case of success and if the cluster consists of more than one
particle, the second particle tries to follow the first one with probability
$\mu$, which is, in general, different from $p$. So do the third,
fourth, etc., particles until either some particle of the cluster
has failed to jump or the cluster has ended. In other words, for $k-$particle
cluster with $k>1$ the following outcomes are possible: 
\begin{enumerate}
\item all $k$ particles stay with probability $(1-p)$; 
\item $m<k$ particles make a step with probability $p\mu^{m-1}\left(1-\mu\right)$; 
\item all $k$ particles jump with probability $p\mu^{k-1}$ 
\end{enumerate}
Then we go to the next cluster in a counterclockwise direction and
continue the update until all clusters on the lattice have been updated.
Note that the result clearly does not depend on what cluster we choose
to start. The clusterwise backward-sequential update, i.e. the condition
of starting from the rightmost particle of a cluster excludes the
situation when the tail of a cluster is updated before its head, which
would occur with the conventional sitewise backward update, when the
sites $1$ and $L$ are inside the same cluster. It is also easy to
see that the exclusion rule is automatically satisfied.

The above formulation uses two control parameters $p$ and $\mu$
having a meaning of probabilities, hence varying in the range $0\leq p,\mu\leq1$.
The particular cases $\mu=0$ and $\mu=p$ correspond to the TASEP
with parallel and backward-sequential updates, respectively. In the
case $\mu=1$ the probability for all particles of a cluster to follow
the first particle is equal to one. Therefore, clusters can only merge
and no dissociation occurs in this limit.

\begin{figure}
\includegraphics[width=1\columnwidth]{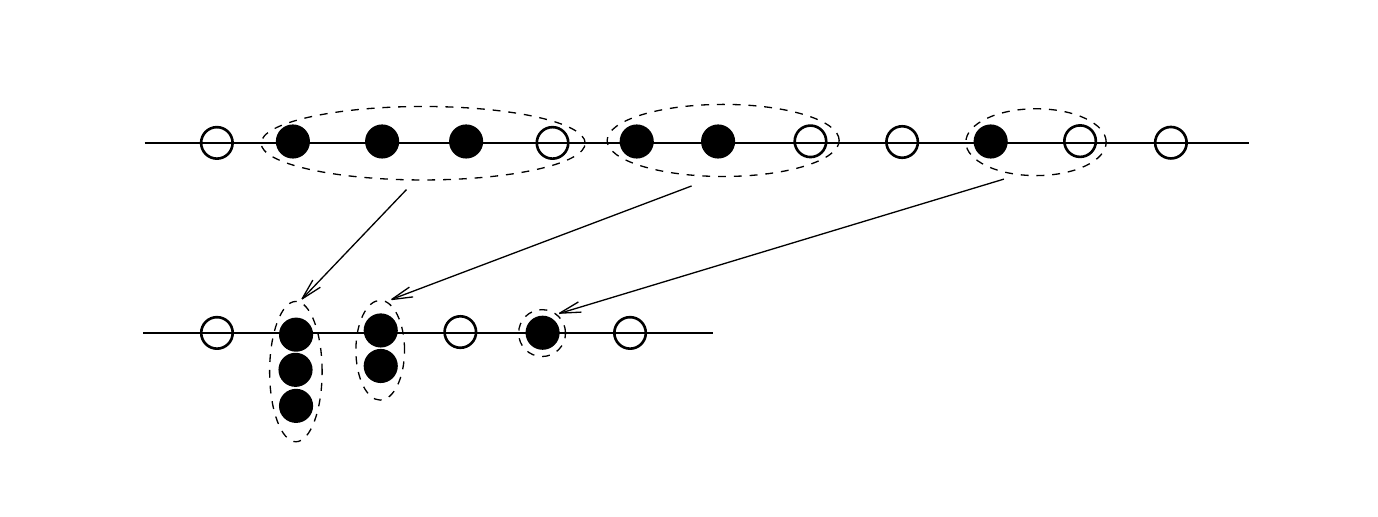}\caption{ZRP-ASEP mapping. }

\label{ASEP-ZRP} 
\end{figure}

Using so-called ZRP-ASEP mapping, the gTASEP can be related to a model
of the ZRP type with an unbounded number of particles in a site. To
establish the correspondence, we replace a string of sites occupied
by a cluster of $n$ particles together with one empty side ahead
by a single site with $n$ particles as shown in Fig. \ref{ASEP-ZRP}.

Thus, $M$ particles are placed on the lattice $\mathcal{L}$ consisting
of $N=L-M$ sites allowed to hold any number of particles unlike at
most one particle per site in the ASEP. The one-step jump of $m$
particles from a splitting cluster will be replaced with a jump of
the same number of particles from the corresponding site to the next
site on the right. According to the above dynamical rules, the jump
of $m$ particles from a site with $n$ particles depends on both
$n$ and $m$, and has the form 
\begin{equation}
\varphi(m|n)=\left\{ \begin{array}{ll}
(1-p), & m=0;\\
p\mu^{m-1}\left(1-\mu\right), & 0<m<n;\\
p\mu^{n-1} & m=n,
\end{array}\right.\label{eq: phi(m|n)}
\end{equation}
for $n>0$ and $\varphi(0|0)=1$ with all the sites being updated
simultaneously at a given time step, as in the parallel update scheme.
As noted in Sec. \ref{sec:Introduction}, the hopping probabilities
of our model are a particular $q=0$ limit of the general three-parametric
hopping probabilities of the model proposed in \cite{chipping}, where
they depended on three parameters $q,\mu$ and $\nu$. Here we use
the notations of \cite{chipping}, which are different from those
used in the first paper on the gTASEP \cite{genTASEP}. Specifically,
the two parameters $\mu$ and $\nu$ of the present article (as well
as of \cite{chipping}) correspond to $p(1+\nu)$ and $\nu p/(1-p)$
of \cite{genTASEP} respectively. The parameter $\nu$ used here is
related to the parameters $p$ and $\mu$ defined above by 
\begin{equation}
\nu=\frac{\mu-p}{1-p}.\label{eq:nu}
\end{equation}
In the following, where it is more appropriate for brevity of notations,
we also use another parameter 
\begin{eqnarray}
\lambda & = & \frac{1}{1-\nu}.\label{eq:lambda}
\end{eqnarray}
In particular, it is convenient for studying the DA limit, which corresponds
to $\lambda\to\infty$.

Note that for periodic lattices, the ZRP-ASEP mapping is not the one-to-one
correspondence between particle configurations, though they can be
made equivalent up to the lattice rotations. In particular, there
are more particle configurations in ASEP-like systems that in ZRP-like
systems. Given two initial particle configurations in the ZRP and
the ASEP related to one another by the mapping, the further processes
are in the one-to-one correspondence in terms of relative distances
between particles, but not in terms of particle coordinates. The essence
of the difference is different translational symmetries of the lattices
with different numbers of sites: A particle configuration returns
to itself after $L$ unit translations in the ASEP and after $M$
unit translations in the ZRP. As seen from the Bethe ansatz solution
below, a little modification is necessary to transform formulas using
the coordinate notations in one system to those in the other. However,
this difference does not affect translation invariant quantities,
e.g. stationary state observables such as particle density and current.
Once calculated in one system such a quantity can be easily related
to similar quantity in the other. The densities of particles defined
in the TASEP-like and and ZRP-like systems as 
\[
c=M/L\,\,\,\mbox{and}\,\,\,\rho=M/N,
\]
respectively, are related by 
\[
c=\frac{\rho}{1+\rho}.
\]
The total numbers of jumps made by all particles are the same in both
systems. In particular, this is the case for the stationary state
average number of jumps per one step $J$. Then the mean velocities
of particles $v=J/M$ are the same in both systems, while the stationary
state currents, i.e., average numbers of particles leaving one site
per time step, $j^{ASEP}=J/L$ and $J^{ZRP}=J/N$, satisfy the following
relation: 
\[
\frac{j^{ASEP}}{c}=\frac{j^{ZRP}}{\rho}=v.
\]
Below, we use these relations to express stationary characteristics
of the gTASEP on terms of quantities obtained for corresponding ZRP-like
system.

\section{Stationary state\label{sec:Stationary-state}}

\subsection{Partition function formalism}

The advantage of the ZRP-like system is that its stationary measure
has a particularly simple form. Specifically, consider the one-dimensional
periodic lattice $\mathcal{L}=\mathbb{Z}/N\mathbb{Z}$ consisting
of $N$ sites with $M$ particles on it. Every site can hold any number
of particles. It is convenient to specify particle configurations
in the ZRP-like systems by $N$ occupation numbers of all sites 
\begin{equation}
\mathbf{n}=\left\{ n_{1},\ldots,n_{N}\right\} .\label{C}
\end{equation}
The system configuration is updated at every time step by bringing
any number $m_{i}\leq n_{i}$ of particles from every site $i=1,\dots,N$
to the next site $i+1$ with probability $\varphi(m_{i}|n_{i})$ defined
in (\ref{eq: phi(m|n)}). Therefore, the probability $P_{t}\left(\mathbf{n}\right)$
for the system to be in a configuration $\mathbf{n}$ at time step
$t$ obeys the Markov equation 
\begin{equation}
P_{t+1}(\mathbf{n})=\sum_{\mathbf{n'}}\mathbf{M_{\mathbf{n,}n'}}P_{t}(\mathbf{n'}),\label{eq:Markov eq.}
\end{equation}
with transition matrix \textbf{M} defined by matrix elements 
\begin{equation}
\mathbf{M_{n,n'}}=\sum_{\{m_{k}\in\mathbb{Z}_{\geq0}\}_{k\in\mathcal{L}}}\prod_{i\in\mathcal{L}}T_{n_{i},n_{i}'}^{m_{i-1},m_{i}},\label{eq:Markov matrix}
\end{equation}
where $T_{n_{i},n_{i}'}^{m_{i-1},m_{i}}=\delta_{(n_{i}-n_{i}'),(m_{_{i-1}}-m_{i})}\varphi(m_{i}|n_{i}')$.
It was proved in \cite{Evans Zia Majumdar} that if and only if the
hopping probabilities $\varphi(m|n)$ have the functional form 
\begin{equation}
\varphi(m|n)=\frac{v(m)w(n-m)}{\sum_{i=0}^{n}v(i)w(n-i)},\label{eq: phi(m|n) - general}
\end{equation}
with two arbitrary positive functions $v(k)$ and $w(k)$, the Markov
equation (\ref{eq:Markov eq.}) has a unique stationary solution,
which belongs to the class of the so-called product measures; i.e.,
the probability of a configuration is given by the product of one-site
factors 
\begin{equation}
P_{st}\left(\mathbf{n}\right)=\frac{1}{Z\left(M,N\right)}\prod_{i=1}^{N}f\left(n_{i}\right),\label{P_st}
\end{equation}
where the one-site factor is given by 
\begin{equation}
f\left(n\right)=\sum_{i=0}^{n}v(i)w(n-i),\label{eq: f(n)}
\end{equation}
and $Z\left(M,N\right)$ is the normalization constant, referred to
as the partition function in statistical physics. In our case, the
functions $v(k)$ and $w(k)$ that define the hopping probabilities
(\ref{eq: phi(m|n)}) can be chosen as 
\begin{eqnarray}
v(k) & = & \mu^{k}(\delta_{k,0}+(1-\delta_{k,0})(1-\nu/\mu)),\label{eq:v(k)}\\
w(k) & = & (\delta_{k,0}+(1-\delta_{k,0})(1-\mu)),\label{eq:w(k)}
\end{eqnarray}
which, according to (\ref{eq: f(n)}), yield an expression for the
single site weight, 
\begin{equation}
f(n)=(\delta_{n,0}+(1-\delta_{n,0})(1-\nu))\label{eq: f(n) form}
\end{equation}

With the product measure (\ref{P_st}) in hands, we are in position
to use the partition function formalism \cite{evans,burda} to calculate
the stationary state observables. The partition function is the normalization
constant of the stationary distribution (\ref{P_st}), given by the
sum of unnormalized weights over all particle configurations, 
\begin{equation}
Z\left(M,N\right)=\sum_{n_{1},\dots,n_{N}\geq0}\delta_{\left\Vert n\right\Vert ,M}\prod_{i=1}^{N}f\left(n_{i}\right),
\end{equation}
Here $f(n)$ is the one-site weight defined in (\ref{eq: f(n) form}),
$\left\Vert n\right\Vert =n_{1}+\dots+n_{N}$, and the Kronecker $\delta$
symbol constrains the summation to particle configurations with the
number of particles fixed to $M$. The sum is given by the contour
integral 
\begin{equation}
Z\left(M,N\right)=\oint_{\Gamma_{0}}\frac{\left[F(z)\right]^{N}}{z^{M+1}}\frac{dz}{2\pi i},\label{Z(N,M)}
\end{equation}
where 
\[
F\left(z\right)=\sum_{n=0}^{\infty}z^{n}f\left(n\right)
\]
is the generating function of the one-site weights, and the contour
of integration is a small circle closed around the point $z=0$ leaving
all other singularities of $F(z)$ outside. The partition function
contains information about the stationary state of the model. In particular,
the probability for a site to be occupied by $n$ particles is 
\begin{equation}
P(n)=f(n)\frac{Z(M-n,N-1)}{Z(M,N)}.\label{eq:P(n)}
\end{equation}

Another correlation function is the probability $\mathcal{H}(k)$
for $k$ particles on the lattice to hop simultaneously, 
\[
\mathcal{H}(k)=\sum_{\mathbf{n,m}\in\mathbb{Z}_{\geq0}^{N}}\delta_{\left\Vert \mathbf{m}\right\Vert ,k}\delta_{\mathbf{\left\Vert n\right\Vert },M}\varphi(\mathbf{m}|\mathbf{n})P_{st}(\mathbf{n}),
\]
where $\varphi(\mathbf{m}|\mathbf{n})=\prod_{1\leq i\leq L}\varphi(m_{i}|n_{i})$
and we suggest that $\varphi(m|n)=0,$ when $m>n$. The corresponding
generating function, 
\begin{equation}
\Psi\left(x\right)\equiv\sum_{n=0}^{M}x^{n}\mathcal{H}(n).\label{Psi(x)}
\end{equation}
can be obtained in the form of contour integral 
\[
\Psi(x)=\frac{1}{Z\left(M,N\right)}\oint_{\Gamma_{0}}\frac{\left[\Phi\left(x,z\right)\right]^{N}}{z^{M+1}}\frac{dz}{2\pi i}
\]
with the two-variable generating function 
\begin{eqnarray*}
\Phi(x,z) & = & \sum_{n=0}^{\infty}\sum_{m=0}^{n}\varphi(m|n)f(n)x^{m}z^{n}.
\end{eqnarray*}
The values of $\mathcal{H}(n)$ can be then represented via contour
integrals with $\Psi(x)$ around $x=0$, while for the moments we
need the derivatives at $x=1$. In particular, the average total number
of particles jumping per unit time can be evaluated as 
\[
J=\Psi'(1).
\]

Using the representation (\ref{eq: phi(m|n) - general}) of $\varphi(m|n)$
and interchanging the order of summations, we obtain 
\[
\Phi(x,z)=V(xz)W(z),
\]
where $V(t)$ and $W(t)$ are the generating functions of the above
sequences $v(k)$ and $w(k)$: 
\begin{eqnarray*}
V(t) & = & \sum_{k=0}^{\infty}v(k)t^{k},\,\,\, W(t)=\sum_{k=0}^{\infty}w(k)t^{k}.
\end{eqnarray*}
 Noticing that $F(z)=V(z)W(z)$, we obtain 
\begin{equation}
J=\frac{N}{Z\left(M,N\right)}\oint_{\Gamma_{0}}\frac{\left[F\left(z\right)\right]^{N}}{z^{M}}\frac{V'(z)}{V(z)}\frac{dz}{2\pi i}.\label{eq:J}
\end{equation}
The generation functions of the sequences $w(k)$, $v(k)$, and $f(k)$
from (\ref{eq:v(k)})--(\ref{eq: f(n) form}) are 
\[
V(t)=\frac{1-\nu t}{1-\mu t},\,\, W(t)=\frac{1-\mu t}{1-t},\,\, F(t)=\frac{1-\nu t}{1-t}.
\]
Then the above integrals can be evaluated in terms of hypergeometric
functions. Specifically, the grand canonical partition function $\left[F(z)\right]^{N}$
and the function $\left[F(z)\right]^{N}V'(z)/V(z)$, are generating
functions of the $_{2}F_{1}$ Gauss hypergeometric functions and $F_{1}$
Appell function respectively. The integrals extract from these series
the coefficients of the terms of order $M$ and $(M-1)$, respectively
(see Appendix \ref{sec:Hypergeometric-functions}). Then, for the
partition function we have 
\[
Z(M,N)=\left(\begin{array}{c}
L-1\\
M
\end{array}\right)\left._{2}F_{1}\right.(-M,-N;1-L;\nu),
\]
while the average number of particles jumping per unit time is 
\begin{equation}
J=\frac{(\mu-\nu)NM}{(L-1)}\frac{F_{1}(1-M;1-N,1;2-L;\nu,\mu)}{\left._{2}F_{1}\right.(-M,-N;1-L;\nu)}.\label{eq:J-1}
\end{equation}
When one of the arguments is zero, the $F_{1}$ Appell function is
reduced to the $_{2}F_{1}$ Gauss function (see appendix \ref{sec:Hypergeometric-functions}).
Therefore, in the limit $\mu=0,$ i.e. the parallel update (PU) case,
we recover the result obtained in \cite{Povolotsky Mendes}: 
\[
J_{PU}=\frac{p}{1-p}\frac{NM}{(L-1)}\frac{\left._{2}F_{1}\right.(1-M;1-N,2-L;-\frac{p}{1-p})}{\left._{2}F_{1}\right.(-M,-N;1-L;-\frac{p}{1-p})}.
\]
The $\nu=0$ case corresponds to the backward-sequential update (BSU),
for which we have a formula 
\[
J_{BSU}=\frac{pNM}{(L-1)}\left._{2}F_{1}\right.(1-M;1,2-L;p),
\]
obtained in \cite{Brankov Papoyan Poghosyan  Priezzhev}. {[}There
is a minor mistype in \cite{Brankov Papoyan Poghosyan  Priezzhev}:
the factor $z$ corresponding to our $p$ is missing from the final
expression, formula (16).{]} Also the $p=1$ limit of (\ref{eq:J-1})
was obtained in \cite{Woelki}. It should be noted also that the $F_{1}$
Appell function is a two-variable reduction of the Lauricella hypergeometric
function $F_{D}$, which depends on an arbitrary number of variables.
The particle current in a particular example of ZRP, where, at most,
one particle may jump from sites with $r\leq K$ particles with arbitrary
probabilities $0<u(r)<1$ and from sites with $r>K$ with probability
$u(r)=1$ was obtained in~\cite{Kanai} in terms of Lauricella hypergeometric
function $F_{D}$ of $K$ variables. Presumably, our case and the
one studied in~\cite{Kanai} can be unified within a larger class
of processes.

The results we have just obtained give the exact formulas for the
partition functions, from which we can also obtain the occupation
number distribution, and the mean particle current on an arbitrary
finite lattice. However, of physical interest is the thermodynamic
limit, 
\begin{equation}
N\to\infty,M\to\infty,M/N=\rho=const\label{eq:thermodynamic}
\end{equation}
It turns out that depending on the scale of the parameter $\lambda,$
two different regimes naturally appear.

\subsection{Asymptotic analysis \label{sub:Asymptotic-analysis}}

\subsubsection*{Saddle point method, $\lambda/N^{2}\to0$}

In the limit (\ref{eq:thermodynamic}), we can try to evaluate the
integrals (\ref{Z(N,M)}) and (\ref{eq:J}) in the saddle point approximation.
The integrals have the form 
\begin{equation}
\mathcal{I}_{N}\left(h(z),g(z)\right)=\oint_{\Gamma_{0}}e^{Nh(z)}g(z)\frac{dz}{2\pi iz},\label{eq:INT}
\end{equation}
where 
\begin{equation}
h(z)=\ln(1-\nu z)-\ln(1-z)-\rho\ln z.\label{eq:h(z)}
\end{equation}
The critical points of the function $h(z)$ are defined by equation
$h'(z)=0,$ which has two solutions 
\begin{equation}
z_{\pm}=1+\frac{(1-\nu)}{2c\nu}\left(1\pm\sqrt{1+\frac{4(1-c)c\nu}{1-\nu}}\right),\label{eq:z_pm}
\end{equation}
where $c=\rho/(1+\rho)$ is the concentration of particles in the
ASEP-like system. The simple analysis shows that $0<z_{-}<1$ and
$z_{+}>1$, $\Re h(z_{+})<0$ and $\Re h(z_{-})>0.$ Therefore 
\[
z=z_{-}
\]
is the point which gives the dominant contribution to the integrals.
We now choose the steepest descent contour being a circle of radius
$z_{-}$with the center at the origin. Then the integral (\ref{eq:INT})
asymptotically is\begin{widetext} 
\begin{equation}
\mathcal{I}_{N}\left(h(z),g(z)\right)=\frac{e^{Nh_{0}}}{\sqrt{2\pi N|h_{2}|}}\left[g_{0}+\frac{1}{2N}\left(\frac{g_{2}}{|h_{2}|}+\frac{g_{1}h_{3}}{h_{2}^{2}}+\frac{g_{0}}{4}\left(\frac{h_{4}}{h_{2}^{2}}+\frac{5h_{3}^{2}}{3|h_{2}|^{3}}\right)\right)+O\left(N^{-2}\right)\right],\label{eq:INT ASYMP}
\end{equation}
\end{widetext}where $g_{k}=\left(\mbox{i}z\partial_{z}\right)^{k}g(z)|_{z=z_{-}}$
and $h_{k}=\left(\mbox{i}z\partial_{z}\right)^{k}h(z)|_{z=z_{-}}.$
Choosing $g(z)=1$ we obtain the leading order of the partition function
\[
Z(M,N)=\mathcal{I}_{N}\left(h(z),1\right)\simeq\frac{\exp\left(Nh_{0}\right)}{\sqrt{2\pi N|h_{2}|}}.
\]
To obtain the occupation number distribution $P(n)$ using (\ref{eq:P(n)})
we need also the value of $Z(M-n,N-1)$, which can be evaluated choosing
$g(z)=z^{n}\exp\left[-h(z)\right].$ As a result, in the limit (\ref{eq:thermodynamic})
for $n$ not too large $P(n)\simeq f(n)g(z_{-})$, from where we have
a usual Gibbs-like form 
\begin{eqnarray*}
P(n) & = & \lambda^{-1}\exp\left(-n/n^{*}-h(z_{-})\right),\,\,\, n>0\\
P(0) & = & \exp\left(-h(z_{-})\right),
\end{eqnarray*}
with $n^{*}=-1/\ln z_{-}$, obtained before in \cite{Woelki}. This
form of the distribution suggests that only sites with finite (mainly
$n\lesssim n^{*}$) occupation numbers have a chance to appear in
a typical configuration in the stationary state.

To interpret this results in terms of the gTASEP, we note that conditioned
to occupied sites, $n>0,$ the distribution obtained gives the cluster
size distribution. It is the geometric distribution $P(l_{\mbox{cl}}=n)=\left(1-z_{-}\right)z_{-}^{n}$
, with the mean cluster size equal to $\left\langle l_{\mbox{cl}}\right\rangle =z_{-}\left(1-z_{-}\right)^{-1}.$
As $\lambda\to\infty,$ the mean cluster size grows as $\left\langle l_{\mbox{cl}}\right\rangle \sim\sqrt{\lambda/\rho}$,
and, correspondingly, the mean number of clusters at the lattice is
$M\left\langle l_{\mbox{cl}}\right\rangle ^{-1}\simeq Lc^{3/2}/\sqrt{\lambda(1-c)}$.
As discussed in the end of this section, the saddle point approximation
is valid as far as $\lambda L^{-2}\to0$. Therefore, these results
hold as far as the number of clusters grows with the system size and
their size is subextensive, i.e., much less than $L$.

To obtain the particle current we must also evaluate the ratio 
\[
J=N\frac{\mathcal{I}_{N}\left(h(z),g(z)\right)}{\mathcal{I}_{N}\left(h(z),1\right)},
\]
with the function $g(z)=zV'(z)/V(z)=\left(1-\mu z\right)^{-1}-\left(1-\nu z\right)^{-1}$.
The next to leading order finite size correction to the particle current
has a universal meaning in context of KPZ theory, which will be discussed
below. Therefore, we keep the terms of the asymptotic expansion up
to the next to leading order, which yields 
\begin{eqnarray}
J & = & Ng_{0}+\frac{1}{2}\left(\frac{g_{1}h_{3}}{\left|h_{2}\right|^{2}}+\frac{g_{2}}{\left|h_{2}\right|}\right)+O(N^{-1}).\label{eq:J asymp}
\end{eqnarray}
From the leading order we obtain the current-density relation (so-called
flow diagram) for the gTASEP, which being expressed in terms of probabilities
$p$ and $\mu,$ reads as 
\begin{eqnarray}
j^{ASEP} & = & \frac{cp(1+(1-2c)\mu)}{2\mu+2c(p(1-\mu)-\mu)}\label{eq:j^TASEP}\\
 & - & \frac{cp\sqrt{(1-\mu)(1-4(1-c)c(p-\mu)-\mu)}}{2\mu+2c(p(1-\mu)-\mu)},\nonumber 
\end{eqnarray}
and reproduces the formula obtained in \cite{Woelki}. The particular
cases of this expression are: 
\begin{description}
\item [{$\mu=0$,}] well known current-density relation for the TASEP with
PU first obtained in studies of traffic models \cite{SSNI}, 
\[
j_{PU}=\frac{1}{2}\left(1-\sqrt{1-4pc(1-c)}\right);
\]

\item [{$\mu=p$,}] BSU \cite{RSSS}, 
\[
j_{BSU}=\frac{(1-c)cp}{1-cp};
\]

\item [{$\mu=1$,}] the limit of DA in which all particles finally stick
together into a single giant cluster, which performs ordinary Bernoulli
random walk, 
\[
j_{DA}=cp.
\]

\end{description}
In the case of PU, the current-density plot is symmetric due to particle-hole
symmetry. The bigger the value of $\mu,$ the more right skewed is
the plot. In the DA limit, it degenerates into the linear function
describing a random walk of a single particle making steps of length
$M$.

The explicit expression of the next to leading order finite-size correction
to the current given in (\ref{eq: current correction (z_)}) of Appendix
\ref{sec:Explicit-expressions-of} is rather cumbersome. However,
it is informative to look at it close to DA limit: 
\begin{eqnarray*}
\mu & \to & 1,\nu\to1,p=\frac{\mu-\nu}{1-\nu}=const.
\end{eqnarray*}
It is convenient to describe this limit in terms of the parameter
$\lambda$ defined in (\ref{eq:lambda}), for which the limit corresponds
to $\lambda\to\infty$. Then the leading asymptotics in $\lambda$
is 
\begin{eqnarray}
j^{ASEP}(L) & - & j^{ASEP}(\infty)\label{eq:current correctioon (asymp)}\\
 & = & \frac{1}{L}\left[\frac{3cp(1-p)}{4\left(1-c\right)}+O\left(\lambda^{-1/2}\right)\right]+O\left(L^{-2}\right).\nonumber 
\end{eqnarray}
Surprisingly the $1/L$ correction saturates to the finite limit as
$\lambda\to\infty$. However, the next orders' corrections diverge
in this limit, so that the effective expansion parameter is $\sqrt{\lambda}/L,$
from where we can estimate the range of validity of the expressions
obtained. One can see that the correction becomes non-neglectable
as soon as $\lambda$ is of the order of $L^{2}.$ In fact, the very
applicability of the saddle point method is violated at this scale.
The reason for that is merging of the saddle points $z_{-}$and $z_{+}$
with the pole of the function under the integral, which makes all
the terms of the expansion of the function $h(z)$ effectively of
the same order. Indeed, though we implicitly assumed that all the
parameters of $h(z)$ are constants as $N$ grows, the saddle point
method can still be applied with $N-$dependent parameters as far
as the limit 
\[
\lim_{N\to\infty}\left|\frac{N^{1-k/2}h_{k}}{h_{2}^{k/2}}\right|=0
\]
holds for $k\geq3$, where $h_{k}$ is the $k$-th derivative of $h(z)$
evaluated at the saddle point $z_{-}$. In the limit $\lambda\to\infty$,
the saddle points are as close to $z=1$ as $z_{\pm}=1\pm\sqrt{1/\rho\lambda}+O(1/\mbox{\ensuremath{\lambda}})$
and $h_{k}$ grows as $h_{k}\sim\lambda^{\frac{k-1}{2}}$. Therefore,
the above limit holds as far as $\lambda N^{-2}\to0$. The situation
when $\lambda N^{-2}\to const>0$, corresponding to the transition
regime, requires a separate asymptotic analysis.

\subsubsection*{Transition regime, $\lambda N^{-2}=const.$}

Let us consider the integral (\ref{Z(N,M)}) representing the partition
function $Z(M,N)$. To evaluate the integral asymptotically we note
that the function under the integral has only two singularities in
the complex plane $z=0$ and $z=1$, and, in particular, is analytic
at infinity. Therefore we can deform the integration contour $\Gamma_{0}$
closed around the origin by a contour $\Gamma_{1}$ closed around
$z=1$:

\begin{equation}
Z(M,N)=-\oint_{\Gamma_{1}}e^{Nh(z)}\frac{dz}{2\pi iz}.\label{intc}
\end{equation}
Apart from specifying position of the contour with respect to singularities,
we can choose it of any form. It is convenient to integrate over a
small circle centered at $z=1$ going close to the saddle points.
Then, instead of the the function $h(z)$, one can use its asymptotic
expansion at this contour. Choosing 
\[
z=1+\frac{e^{\mbox{i}\varphi}}{\sqrt{\rho\lambda}}
\]
we get 
\[
h\left(z\right)=-2\sqrt{\frac{\rho}{\lambda}}\cos\varphi+O(1/\lambda)
\]
and for the integral we obtain 
\begin{eqnarray*}
Z(M,N) & = & \frac{-1}{\sqrt{\rho\lambda}}\int_{0}^{2\pi}e^{-2N\sqrt{\rho/\lambda}\cos\varphi+\mbox{i}\varphi}\frac{d\varphi}{2\pi}+O(N^{-2})\\
 & \simeq & \frac{\theta}{2M}I_{1}(\theta)
\end{eqnarray*}
where $I_{k}(y)$ is the modified Bessel function of the first kind
(for the definition and properties see appendix \ref{sec:Modified-Bessel-function})
and we introduced the scaling parameter 
\begin{equation}
\theta=2N\sqrt{\frac{\rho}{\lambda}},\label{eq:theta}
\end{equation}
which is finite in the limit under consideration and is supposed to
control the transition from the KPZ to the DA regime. For the occupation
number probability distribution $P(n)$ we also need $Z(M-n,N-1)$,
which in the leading order is obtained from the above equation by
replacing: $M\to M-n,$ $\rho\to\rho-n/N$ and $\theta\to\theta\sqrt{1-n/M}$.
Then for the occupation number probability distribution we have 
\begin{eqnarray}
P(0) & \simeq & 1-\frac{\theta}{2N}\frac{I_{0}(\theta)}{I_{1}(\theta)},\label{cluster1}\\
P\left(n\right) & \simeq & \frac{\theta^{2}}{4NM}\frac{I_{1}\left(\theta\sqrt{1-\frac{n}{M}}\right)}{I_{1}\left(\theta\right)\sqrt{1-\frac{n}{M}}},\,\,\,0<n<M,\label{eq: cluster2}\\
P(M) & \simeq & \frac{\theta}{2N}\frac{1}{I_{1}(\theta)}.\label{eq:cluster3}
\end{eqnarray}
Here we kept only the leading order of the expansion of $P(n)$ for
$n>0$ and two highest orders in $P(0)$ (the latter can actually
be obtained from the former by normalization). It is clear from the
first line that the typical configuration contains only a finite number
of sites are occupied. Consider now the distribution of the random
variable $\chi=n/M\in[0,1]$ conditioned at $n>0,$ i.e. only the
occupied sites are counted. To evaluate the conditional probability,
we divide the right-hand side of (\ref{eq: cluster2}) and (\ref{eq:cluster3})
by $\mathrm{Prob}(n>0)=(1-P(0))$. The distribution obtained has well
defined limiting behavior as $N\to\infty$ being parametrized by single
parameter $\theta:$ 
\begin{eqnarray}
\mbox{Prob}(\chi=1) & = & \frac{1}{I_{0}(\theta)},\label{eq:Prob(n/M=00003D1|n>0)}\\
\mbox{Prob}(\chi<x) & = & \frac{\theta}{2I_{0}\left(\theta\right)}\int_{0}^{x}\frac{I_{1}\left(\theta\sqrt{1-y}\right)}{\sqrt{1-y}}dy.\label{eq:Prob(n/M<x|n>0)}
\end{eqnarray}
A finite fraction of the distribution is concentrated at single point
$\chi=1$ and the rest is the continuous distribution on $[0,1)$.
In terms of the gTASEP the probability (\ref{eq:Prob(n/M=00003D1|n>0)})
is exactly the limiting fraction of time, which all particles spend
in a single cluster. The rest of the time finitely many clusters of
macroscopic size, $n\sim M$, exist on a lattice. For $\chi<1$ the
distribution of the fraction of all particles contained in a given
cluster converges to the continuous distribution (\ref{eq:Prob(n/M<x|n>0)}).
The mean size of a cluster is $\left\langle l_{\mbox{cl}}\right\rangle \simeq2MI_{1}(\theta)\left[\theta I_{0}(\theta)\right]^{-1}$
and, $M/\left\langle l_{\mbox{cl}}\right\rangle $ is the expected
number of clusters, which starting from one at $\theta=0$ approaches
a linear growth $M/\left\langle l_{\mbox{cl}}\right\rangle \simeq\theta/2$
as $\theta\to\infty.$

\subsection{Transfer matrix approach and correlation length\label{sub:Transfer-matrix-approach}}

In addition to the above calculations with the ZRP-like system, which
is suitable for characterizing the current and cluster size distribution,
one can look at the system directly in the ASEP formulation, which
is more appropriate for for study of the stationary correlation functions.
The stationary measure of particle configurations in the TASEP-like
system is similar to that of the ZRP-like system up to the symmetry
with respect to lattice rotation. Specifically, we must replace every
site occupied with $n$ particles with a cluster of $n$ particles
plus one empty site. Looking at formulas for the stationary measure
of a ZRP-like system (\ref{P_st}) and (\ref{eq: f(n) form}) we assign
the weight $(1-\nu)$ to each cluster and the weight $1$ to each
empty site. It is convenient to study the system in the grand canonical
ensemble, where in addition to the above cluster weights we attach
the fugacity $z$ to each particle. Then, the stationary probability
of a particle configuration $\bm{\tau}=(\tau_{1},\dots,\tau_{L}),$
with occupation numbers $\tau_{1},\dots,\tau_{L}=0,1$, will be 
\[
P_{st}(\bm{\tau})=\frac{1}{\mathcal{Z}_{L}(z)}T_{\tau_{1},\tau_{2}}\dots T_{\tau_{L-1},\tau_{L}}T_{\tau_{L},\tau_{1}},
\]
where $T_{0,0}=1$, $T_{0,1}=T_{1,0}=\sqrt{z(1-\nu)},$ and $T_{1,1}=z$.
This measure is similar to the Gibbs measure of the 1D Ising model,
as was first observed in \cite{SSNI} in the context of the TASEP
with PU. Correspondingly, for periodic boundary conditions the partition
function is given by the trace of $L$-th power of the transfer matrix
\[
\mathcal{Z}_{L}(z)=\mbox{Tr}T^{L}=\lambda_{1}^{L}+\lambda_{2}^{L},
\]
where 
\begin{eqnarray*}
\lambda_{1} & = & \frac{1}{2}\left(1+z+\sqrt{(z+1)^{2}-4\nu z}\right),\\
\lambda_{2} & = & \frac{1}{2}\left(1+z-\sqrt{(z+1)^{2}-4\nu z}\right),
\end{eqnarray*}
are eigenvalues of the matrix $T$, defined so that $\lambda_{1}>\lambda_{2}\geq0$.
The largest eigenvalue $\lambda_{1}$ defines the specific free energy,
\[
f(z)=-\lim_{L\to\infty}\frac{\ln\mathcal{Z}_{L}(z)}{L}=-\ln\lambda_{1}.
\]
The density of particles is fixed by the thermodynamic relation 
\begin{equation}
c=-z\partial_{z}f(z).\label{eq:density-fugacity}
\end{equation}
This is the quadratic equation for $z$ with two roots, which interchange
under replacement $c\longleftrightarrow1-c$. Which one to choose
is to be decided from direct evaluation of the particle density, i.e.
of one-point correlation function. In general to evaluate $s-$point
correlation functions of the form $\left\langle \tau_{k_{1}}\dots\tau_{k_{s}}\right\rangle $,
where $\left\langle a\right\rangle $ is the notation for expectation
value of the random variable $a$, one has to insert the matrix 
\[
\widehat{\tau}=\left(\begin{array}{cc}
0 & 0\\
0 & 1
\end{array}\right)
\]
into the product of transfer matrices in the places corresponding
to sites $k_{1},\dots,k_{s}$: 
\[
\left\langle \tau_{k_{1}}\dots\tau_{k_{s}}\right\rangle =\frac{\mbox{Tr}\left[T^{k_{1}}\widehat{\tau}T^{k_{1}+k_{2}}\widehat{\tau}\dots\widehat{\tau}T^{L-(k_{1}+\dots+k_{s})}\right]}{\mathcal{Z}_{L}(z)}.
\]
To evaluate expressions of this kind we also need the eigenvectors
of $T$, 
\begin{eqnarray*}
\mathbf{v}_{1} & = & \left[\begin{array}{c}
\left(\frac{\sqrt{(z+1)^{2}-4\nu z}-z+1}{2\sqrt{(z+1)^{2}-4\nu z}}\right)^{\frac{1}{2}}\\
\\
\left(\frac{\sqrt{(z+1)^{2}-4\nu z}+z-1}{2\sqrt{(z+1)^{2}-4\nu z}}\right)^{\frac{1}{2}}
\end{array}\right],\\
\end{eqnarray*}
and 
\begin{eqnarray*}
\mathbf{v}_{2} & = & \left[\begin{array}{c}
-\left(\frac{\sqrt{(z+1)^{2}-4\nu z}+z-1}{2\sqrt{(z+1)^{2}-4\nu z}}\right)^{\frac{1}{2}}\\
\\
\left(\frac{\sqrt{(z+1)^{2}-4\nu z}-z+1}{2\sqrt{(z+1)^{2}-4\nu z}}\right)^{\frac{1}{2}}
\end{array}\right],\\
\end{eqnarray*}
corresponding to $\lambda_{1}$ and $\lambda_{2}$ respectively, which
are normalized to $\left\Vert v_{1}\right\Vert =\left\Vert v_{2}\right\Vert =1.$
Verifying the identity $c=\left\langle \tau\right\rangle =\mbox{Tr}(\widehat{\tau}T^{L})\simeq(\mathbf{v}_{1},\widehat{\tau}\mathbf{v}_{1}),$
we see that the root of (\ref{eq:density-fugacity}) to be chosen
is 
\begin{eqnarray}
\!\!\!\!\!\! z^{*} & = & 1-2\left(1+\frac{\sqrt{(1-\nu)\left(1-\nu(1-2c)^{2}\right)}}{(1-2c)(1-\nu)}\right)^{\!\!-1}\!\!\!\!\!\!.\label{eq:z^*}
\end{eqnarray}
With the use of relations $(\mathbf{v}_{1},\widehat{\tau}\mathbf{v}_{1})=c,(\mathbf{v}_{2},\widehat{\tau}\mathbf{v}_{2})=(1-c),$
and $(\mathbf{v}_{1},\widehat{\tau}\mathbf{v}_{2})=\sqrt{c(1-c)}$
the two point correlation function is given by \begin{widetext} 
\begin{eqnarray*}
\left\langle \tau_{1}\tau_{1+k}\right\rangle  & = & \frac{c^{2}+(1-c)^{2}e^{-L/\xi}+c(1-c)(e^{-k/\xi}+e^{-(L-k)/\xi})}{1+e^{-L/\xi}}
\end{eqnarray*}
\end{widetext} where 
\begin{equation}
\xi\equiv\frac{1}{\ln(\lambda_{1}/\lambda_{2})}=-\left[\ln\left(1-\frac{2}{1+\sqrt{\frac{1-(1-2c)^{2}\nu}{1-\nu}}}\right)\right]^{-1}\label{eq:corr length}
\end{equation}
is the correlation length. When the correlation length is finite,
i.e. for $\nu<1,$ the two-point correlator becomes a product of one
point correlators $\left\langle \tau_{1}\tau_{1+k}\right\rangle \to c^{2}$
at large distances, $L\gg k\to\infty.$ As expected, the covariance
decays exponentially in the range $\xi\ll k\ll L$, 
\begin{equation}
C(k)\equiv\left\langle \tau_{1}\tau_{1+k}\right\rangle -\left\langle \tau_{1}\right\rangle \left\langle \tau_{1+k}\right\rangle \simeq c(1-c)e^{-k/\xi},\label{eq:covariance}
\end{equation}
which justifies $\xi$ being the correlation length. As $\nu\to1$,
i.e., $\lambda\to\infty$, the correlation length diverges as 
\[
\xi\simeq\sqrt{\lambda c(1-c)}.
\]
In the transition regime, $\lambda\sim L^{2},$ the correlation range
is of the order of the system size $L$. Expressed in terms of the
distance $r$ measured in units of the system size the covariance
has the form 
\[
C(Lr)=\frac{(1-2c)e^{-1/\widetilde{\xi}}+c(1-c)(e^{-r/\widetilde{\xi}}+e^{-(1-r)/\widetilde{\xi}})}{1+e^{-1/\widetilde{\xi}}},
\]
where $\widetilde{\xi}=2c(1-c)/\theta$ is the effective correlation
length in the system size scale, which depends on the transition parameter
$\theta$ defined in (\ref{eq:theta}).

\section{Statistics of particle current\label{sec:Statistics-of-particle}}

\subsection{Bethe ansatz and method by Derrida-Lebowitz}

To characterize current fluctuations we introduce a deformed Markov
matrix $\mathbf{M}^{\gamma}$, depending on an auxiliary parameter
$\gamma$, where every particle step is supplied with an extra weight
$\exp\gamma$. Then the matrix elements are defined as follows 
\[
\mathbf{M}_{\mathbf{n,n'}}^{\gamma}=\mathbf{M}_{\mathbf{n,n'}}\exp\left(\gamma\mathcal{N}(\mathbf{n,n'})\right),
\]
where $\mathbf{M}_{\mathbf{n,n'}}$ --- are matrix elements of the
original Markov matrix and $\mathcal{N}(\mathbf{n,n'})$ is the number
of particle jumps in the one-step transition from $\mathbf{n}'$ to
$\mathbf{n}$. This matrix governs the evolution of the generating
function $G_{t}\left(\mathbf{n,\gamma}\right)=\sum_{Y=0}^{\infty}e^{\gamma Y}P_{t}\left(\mathbf{n},Y\right),$
of the joint probability $P_{t}\left(\mathbf{n},Y\right)$ for the
system to be in configuration $\mathbf{n}$ at time $t$, while the
total distance $Y_{t}$ traveled by all particles is equal to $Y$,
\[
\mathbf{G}_{t+1}=\mathbf{M^{\gamma}}\mathbf{G}_{t},
\]
where $\mathbf{G}_{t}$ is the column vector with components $G_{t}\left(\mathbf{n,\gamma}\right)$.
The generating function $\left\langle e^{\gamma Y_{t}}\right\rangle $
of moments of $Y_{t}$ is a sum of $\sum_{\mathbf{n}}G_{t}\left(\mathbf{n,\gamma}\right)$
over all configurations. Its large time behavior is dominated by the
eigenvalue $\Lambda_{0}\left(\gamma\right)$ of the matrix $\mathbf{M^{\gamma}}$
with the largest real part, and, hence, the logarithm of $\Lambda_{0}\left(\gamma\right)$
is the scaled generating function of cumulants of $Y_{t}$, 
\begin{equation}
\ln\Lambda_{0}\left(\gamma\right)=\lim_{t\rightarrow\infty}\frac{\ln\left\langle e^{\gamma Y_{t}}\right\rangle }{t}.\label{comulant}
\end{equation}

The diagonalization of matrix $\mathbf{M^{\gamma}}$ using the Bethe
ansatz technique was described in details in \cite{genTASEP,chipping}
for both ASEP and ZRP-like versions of our system. Let us first briefly
discuss the latter. Alternatively to the set of occupation numbers
$\mathbf{n}$, it is convenient to describe particle configurations
in terms of coordinates of particles on the lattice $\mathcal{L}$,
\[
\mathbf{y}=(1\leq y_{1}\leq\dots\leq y_{M}\leq N)
\]
in the same way as we do for the ASEP, except that the coordinates
are weakly ordered, because many particles in a site are allowed.
Then, the components of eigenvectors $\mathbf{\Psi}$ of $\mathbf{M^{\gamma}}$
are looked for in the form 
\begin{equation}
\mathbf{\Psi_{n}}(\mathbf{z})=P_{st}\left(\mathbf{n}\right)\Psi^{0}(\mathbf{y}|\mathbf{z}),\label{eq: eigenvector}
\end{equation}
where $P_{st}\left(\mathbf{n}\right)$ is the stationary state weight
of particle configuration, and 
\begin{equation}
\Psi^{0}(\mathbf{y}|\mathbf{z})=\sum_{\sigma\in S_{M}}A_{\sigma}z_{\sigma_{1}}^{y_{1}}\dots z_{\sigma_{M}}^{y_{M}}\label{eq: Bethe ansatz}
\end{equation}
is the Bethe function depending on $M$-tuple $\mathbf{z}=(z_{1},\dots,z_{M})$
of complex numbers to be defined later. Here, $\mathbf{y}$ are weakly
increasing coordinates of particles corresponding to the occupation
numbers $\mathbf{n}$, the summation is over the permutations $\sigma=(\sigma_{1},\dots,\sigma_{M})$
of the numbers $(1,\dots,M)$, and $A_{\sigma}$ are the permutation-dependent
coefficients defined by relation 
\begin{equation}
\frac{A\dots ij\dots}{A\dots ji\dots}=-\frac{\Big(1-e^{-\gamma}z_{i}\Big)\Big(\nu-e^{-\gamma}z_{j}\Big)}{\Big(1-e^{-\gamma}z_{j}\Big)\Big(\nu-e^{-\gamma}z_{i}\Big)}.\label{eq:A_ij/A_ji}
\end{equation}
One can show that action of the matrix $\mathbf{M^{\gamma}}$ on $\mathbf{\Psi}$
is reduced to multiplication by the eigenvalue 
\begin{equation}
\Lambda(\gamma)=\prod_{i=1}^{M}\Big(e^{\gamma}pz_{i}^{-1}+(1-p)\Big),\label{eq:eigenvalue}
\end{equation}
provided that the Bethe function satisfies periodic boundary conditions
$\Psi^{0}(y_{1},\dots,y_{M}|\mathbf{z})=\Psi^{0}(y_{2},\dots,y_{M},y_{1}+N|\mathbf{z})$,
which are equivalent to the system of $M$ algebraic Bethe ansatz
equations (BAEs), 
\begin{equation}
z_{i}^{N}=\left(-1\right)^{M-1}\prod_{j=1}^{M}\frac{\Big(1-e^{-\gamma}z_{i}\Big)\Big(\nu-e^{-\gamma}z_{j}\Big)}{\Big(1-e^{-\gamma}z_{j}\Big)\Big(\nu-e^{-\gamma}z_{i}\Big)}.\label{eq:BAE ZRP}
\end{equation}
for the numbers $z_{1},\dots,z_{M}$.

Though the above analysis is applied to the ZRP-like system, the minor
modification is required to the ASEP. The eigenvector of the corresponding
Markov matrix is that in (\ref{eq: Bethe ansatz}), except that the
particle coordinates are read off in a different way. Specifically,
the ASEP coordinates $\mathbf{x}$ are obtained from the ZRP coordinates
by a shift 
\[
(x_{1},x_{2},\dots,x_{M})=(y_{1},y_{2}+1\dots,y_{M}+M-1),
\]
which ensures them to be strictly increasing as necessary. One can
also look for the eigenvector right in the form (\ref{eq: Bethe ansatz})
in terms of the ASEP coordinates $\mathbf{x}$. Then we will have
to multiply the ratio of amplitudes (\ref{eq:A_ij/A_ji}) by the factor
$z_{i}/z_{j}$. The form (\ref{eq:eigenvalue}) of the eigenvalues
stays the same and the periodic boundary conditions $\Psi^{0}(x_{1},\dots,x_{N}|\mathbf{z})=\Psi^{0}(x_{2},\dots,x_{N},x_{1}+L|\mathbf{z})$
yield the BAE 
\begin{equation}
z_{i}^{L}=\left(-1\right)^{M-1}\prod_{j=1}^{M}\frac{z_{i}\Big(1-e^{-\gamma}z_{i}\Big)\Big(\nu-e^{-\gamma}z_{j}\Big)}{z_{j}\Big(1-e^{-\gamma}z_{j}\Big)\Big(\nu-e^{-\gamma}z_{i}\Big)},\label{eq:BAE TASEP}
\end{equation}
which are different from (\ref{eq:BAE ZRP}) in a single factor $\prod_{j=1}^{M}z_{j}$.

The problem of finding the largest eigenvalue for both models is reduced
to identifying a particular solution of the BAE corresponding to the
ground state. To this end, we note that in the limit $\gamma\to0$
the matrix $\mathbf{M}^{\gamma}$ turns to the transition probability
matrix $\mathbf{M}$ having the largest eigenvalue equal to one, so
that we expect $\Lambda_{0}(\gamma)\to1$ as $\gamma\to0$. In addition,
the corresponding eigenvector becomes the stationary state in this
limit, which can be obtained from (\ref{eq: eigenvector}) and (\ref{eq: Bethe ansatz})
by setting $z_{1}=\dots=z_{M}=1$. Taking the product of all $M$
equations in both systems, we see that all solutions satisfy the constraints
$\left(\prod_{j=1}^{M}z_{j}\right)^{N}=1$ and $\left(\prod_{j=1}^{M}z_{j}\right)^{L}=1$
for (\ref{eq:BAE ZRP}) and (\ref{eq:BAE TASEP}), respectively. Therefore,
the sets of solutions of BAE can be classified into sectors, where
the product of Bethe roots equals different roots of unity independent
of $\gamma$. In particular, continuing the $\gamma=0$ limit to arbitrary
values of $\gamma,$ we see that in both systems the ground states
belong to the sector

\begin{equation}
\prod_{j=1}^{M}z_{j}=1,\label{eq:translation invariance}
\end{equation}
where the systems (\ref{eq:BAE ZRP}) and (\ref{eq:BAE TASEP}) are
identical and we can use either of them to obtain the eigenvalue $\Lambda_{0}(\gamma)$.
The product of the Bethe roots is the factor that the Bethe eigenfunction
multiplies by under the unit translation, i.e. the eigenvalue of the
translation operator commuting with the matrix $\mathbf{M}^{\gamma}$.
Its value reflects the translation invariance of the ground state
mentioned before.

To find the solution of (\ref{eq:BAE ZRP}), we first make a change
of variables, 
\begin{equation}
z_{i}=e^{\gamma}\frac{1-\nu u_{i}}{1-u_{i}}.
\end{equation}
In the variables $u_{i}$, the BAE and the eigenvalue $\Lambda(\gamma)$
simplify to the following form 
\begin{eqnarray}
\left(\frac{1-\nu u_{i}}{1-u_{i}}\right)^{N}e^{N\gamma} & = & (-1)^{M-1}\prod_{j=1}^{M}\frac{u_{i}}{u_{j}},\label{BAEy}\\
\Lambda\left(\gamma\right) & = & \prod_{i=1}^{M}\left(\frac{1-\mu u_{i}}{1-\nu u_{i}}\right).\label{Lambda(gamma)y}
\end{eqnarray}
The method used by Derrida and Lebowitz to find eigenvalues for the
TASEP is based on the observation that the solutions of BAE can be
found among the roots of a single polynomial, 
\[
\mathcal{P}(u)=\left(1-\nu u\right)^{N}B-(1-u)^{N}u^{M},
\]
where $B=\left(-1\right)^{M-1}e^{\gamma N}\prod_{j=1}^{M}u_{j}$ is
the parameter that itself is a function of the solution. What we actually
need is to evaluate the sums of values that particular functions take
on the roots from the solution of interest. These sums can be evaluated
using the Cauchy theorem, 
\begin{equation}
\sum_{i=1}^{M}f(u_{j})=\oint_{\Gamma_{0}}f(u)\frac{\mathcal{P}'(u)}{\mathcal{P}(u)}\frac{du}{2\pi\mbox{i}},\label{eq:sum over roots}
\end{equation}
where the integration is over the contour enclosing all the necessary
roots and the function $f(u)$ is analytic inside the contour. In
our case the roots from the solution corresponding to the ground state
are those $M$ roots $z_{1},\dots,z_{M}$, which approach one as $\gamma\to0$
or zero in terms of the variables $u_{1},\dots,u_{M}.$ Choosing the
function $f(u)$ in the form $f(u)=\ln\left[(1-\mu u)/(1-\nu u)\right]$
we obtain after the integration by parts the logarithm of the largest
eigenvalue as function of $B$: 
\begin{eqnarray}
\ln\Lambda_{0}(\gamma) & =\left(\mu-\nu\right) & \oint_{\Gamma_{0}}\frac{\ln\left[1-\frac{B(1-\nu u)^{N}}{(1-u)^{N}u^{M}}\right]}{\left(1-\mu u\right)(1-\nu u)}\frac{du}{2\pi i}.\label{Labda(B)}
\end{eqnarray}
Here the integration is over the contour satisfying to the condition
$|B(1-\nu u)^{N}/(1-u)^{N}u^{M}|<1$ and enclosing $M$ roots of $\mathcal{P}(u)$
located near the origin. Note that the contour does not cross any
branch cuts of the logarithm, which can be chosen connecting $M$
roots inside the contour to the origin and the other $N$ roots outside
the contour to $u=1.$ Such a contour exists if $|B|$ is small enough.
The relation of $\gamma$ and $B$ can be recovered from the translation
invariance condition (\ref{eq:translation invariance}), which after
taking a logarithm and going to the variables $u_{i}$ yields 
\begin{equation}
\gamma=\frac{1-\nu}{M}\oint_{\Gamma_{0}}\frac{\ln\Bigg(1-\frac{B(1-\nu u)^{N}}{(1-u)^{N}u^{M}}\Bigg)}{\left(1-u\right)\left(1-\nu u\right)}\frac{du}{2\pi i}.\label{gamma(B)}
\end{equation}
Note also, that what we actually integrated by parts to arrive at
formulas (\ref{Labda(B)}) and (\ref{gamma(B)}) were not exactly
the original expressions given by (\ref{eq:sum over roots}), but
the ones obtained by addition of terms analytic inside the contour
of integration, which, hence, are integrated out to zero. 

To evaluate the integrals we use a series expansion of the logarithms
in powers of $B$ and integrate the resulting series term by term.
Integrations can explicitly be performed in terms of the Appell and
Gauss hypergeometric functions:

\begin{widetext} 
\begin{eqnarray}
\ln\Lambda_{0}(\gamma) & = & -(\mu-\nu)\sum_{n=1}^{\infty}\frac{B^{n}}{n}\binom{Ln-2}{Mn-1}F_{1}\left(1-nM;1-nN,1;2-nL;\nu,\mu\right),\label{eq:Lambda-ser}\\
\gamma & = & -\frac{1-\nu}{M}\sum_{n=1}^{\infty}\frac{B^{n}}{n}\binom{Ln-1}{Mn-1}\left._{2}F_{1}\right.\left(\begin{array}{c}
1-Mn,1-Nn\end{array};1-nL;\nu\right).\label{eq:gamma-ser}
\end{eqnarray}

\end{widetext}The scaled cumulants 
\[
c_{n}\equiv\lim_{t\to\infty}\left.\frac{\partial^{n}}{t\partial\gamma^{n}}\left\langle e^{\gamma Y_{t}}\right\rangle \right|_{\gamma=0}=\left.\frac{\partial^{n}}{\partial\gamma^{n}}\ln\Lambda_{0}(\gamma)\right|_{\gamma=0}.
\]
of the particle current $Y_{t}$ can be obtained as the coefficients
of the power expansion of $\ln\Lambda_{0}(\gamma)$ in $\gamma$,
which can be constructed to any finite order by eliminating $B$ between
the two series. In particular, using the Euler transformation for
hypergeometric functions (\ref{eq:Euler}) the first scaled cumulant
$c_{1}=\lim_{t\to\infty}t^{-1}\left\langle Y_{t}\right\rangle $,
the number of particle jumps per unit time, can be shown to coincide
with $J$ from (\ref{eq:J-1}) obtained by averaging over the stationary
state. The second cumulant, the scaled variance of $Y_{t}$, is related
to the diffusion coefficient for a particle $\Delta=M^{-2}c_{2}$.
The exact value of the latter is

\begin{widetext}

\begin{eqnarray*}
\Delta & =\lambda p\frac{\binom{2L-2}{2M-1}}{\binom{L-1}{M-1}^{2}} & \left[\frac{(2L-1)}{2(L-1)}\frac{F_{1}(1-M;1-N,1;2-L;\nu,\mu)\left._{2}F_{1}\right.\left(1-2M,1-2N,1-2L;\nu\right)}{\left[\left._{2}F_{1}\right.\left(1-M,1-N,1-L;\nu\right)\right]^{3}}\right.\\
 &  & \left.\hspace{60mm}-\frac{F_{1}(1-2M;1-2N,1;2-2L;\nu,\mu)}{\left[\left._{2}F_{1}\right.\left(1-M,1-N,1-L;\nu\right)\right]^{2}}\right],
\end{eqnarray*}

\end{widetext} from which, using the identities for Gauss and Appell
functions we can recover the corresponding quantities for particular
cases of PU, $\mu=0$, BSU, $\nu=0$ and the DA limit $\mu\to\nu=1,p=const$.
The exact formula of cumulant $c_{n}$ is already rather cumbersome
for $n=2$, and it becomes more and more complicated as $n$ grows.
Of major interest is the scaling behavior of the cumulants and the
whole function $\Lambda_{0}(\gamma),$ which is also related to the
LDF of particle current.

\subsection{Scaling limits \label{sub:Scaling-limits}}

We would like to investigate the thermodynamic limit 
\[
M,N\to\infty,M/N=\rho.
\]
The structure of terms of the series obtained for $\Lambda_{0}(\gamma)$
and $\gamma$ is very similar to that of the integrals analyzed above
for the partition function and average current. Thus, the same asymptotic
analysis is applicable. Again, depending on the scale of $\lambda$
there are two different regimes: the first, where the integrals can
be analyzed in the saddle point approximation and the second where
the integrals can be evaluated in terms of modified Bessel functions.
It is worth emphasizing again that the saddle point approximation
is valid at any scale of $\lambda$ satisfying $\lambda N^{-2}\to0.$
Thus, the universal KPZ scaling function obtained in this approximation
holds through a range of scales, with the scale entering only to the
non-universal scaling constants. Then, the as the parameter $\lambda N^{-2}$
varies from zero to infinity, the KPZ-Gauss transition takes place.

\subsubsection{KPZ regime, $\lambda N^{-2}\to0,\gamma\to0,\gamma\lambda^{1/4}N^{3/2}=const,$}

Up to the $1/n$ factor and the common factors before the integrals
the terms of the order $n$ of series (\ref{eq:Lambda-ser}) and (\ref{eq:gamma-ser})
are given by the integrals $\mathcal{I}_{nN}(h(z),g(z))$ of the form
(\ref{eq:INT}) with the function $h(z)$ defined in (\ref{eq:h(z)})
and instead of the function $g(z)$ we substitute $r(z)=z\left[\left(1-\mu z\right)\left(1-\nu z\right)\right]^{-1}$
for $\Lambda_{0}(\gamma)$ series and $s(z)=z\left[\left(1-z\right)\left(1-\nu z\right)\right]^{-1}$
for the terms of $\gamma$ series. To obtain the meaningful precision
it is enough to keep only leading order terms in asymptotics of $\gamma$
and the next to the leading order terms for the eigenvalue. Evaluating
$\mathcal{I}_{nN}(h(z),r(z))$ and $\mathcal{I}_{nN}(h(z),s(z))$
with the help of (\ref{eq:INT ASYMP}) in two leading orders we obtain
the universal scaling form obtained first in \cite{Derrida Lebowitz},
\begin{equation}
\ln\Lambda(\gamma)=J_{\infty}\gamma+aN^{-3/2}G(bN^{3/2}\gamma),\label{eq:ldf-kpz}
\end{equation}
where $G(z)$ has a parametric representation 
\begin{eqnarray}
G(z) & = & -\mbox{Li}_{5/2}(t),\,\,\, z=-\mbox{Li}_{3/2}(t),\label{eq:Der-Lib}
\end{eqnarray}
via the polylogarithm function $\mbox{Li}_{s}(x)=\sum_{i>0}x^{i}/i^{s}$.
The infinite volume current 
\[
J_{\infty}=Mpr_{0}/s_{0}=Lj^{ASEP}
\]
coincides with the particle current obtained from the averaging over
stationary state, the coefficients $a$ and $b,$ 
\[
a=\frac{\mu-\nu}{2\sqrt{2\pi|h_{2}|}}\left(\frac{r_{2}-s_{2}/s_{0}}{|h_{2}|}+\frac{(r_{1}-s_{1}/s_{0})h_{3}}{h_{2}^{2}}\right)
\]
and 
\[
b=\frac{\sqrt{2\pi|h_{2}|}}{\rho s_{0}(1-\nu)}
\]
are the non-universal model-dependent constants expressed via the
coefficients of expansion of the functions $h(z),r(z),s(z)$ in the
dominant saddle point $z_{-}$: $r_{k}=\left(\mbox{i}z\partial_{z}\right)^{k}r(z)|_{z=z_{-}}$,
$s_{k}=\left(\mbox{i}z\partial_{z}\right)^{k}s(z)|_{z=z_{-}}$ and
$h_{k}=\left(\mbox{i}z\partial_{z}\right)^{k}h(z)|_{z=z_{-}}$ . The
explicit expressions of these constants can be found in Appendix \ref{sec:Explicit-expressions-of}
.

It is clear from (\ref{eq:ldf-kpz}) that nontrivial scaling occurs,
when $\gamma$ is of order of $b^{-1}N^{-3/2},$ which is of order
of $\lambda^{-1/4}N^{-3/2}$ as $\lambda\to\infty.$ The scaling form
(\ref{eq:ldf-kpz}) suggests that at large time deviations of the
time-averaged current from its thermodynamic value are of the form
\begin{equation}
\mbox{Prob}\left(Y_{t}/t<y\right)\sim\exp\left[-atN^{-3/2}\widehat{G}\left(\frac{y-J_{\infty}}{ab}\right)\right],\label{eq:ldf}
\end{equation}
where the scaling function $\widehat{G}(x)=\sup_{t}\left(xt-G(t)\right)$
is a Legendre transform of function $G(t)$. Appearance of the factor
$tN^{-3/2}$ is the universal KPZ-specific feature, which is akin
to the fact that the dynamical exponent of the KPZ class is $z=3/2.$
Specifically, the term $atN^{-3/2}$ is  the parameter, which is supposed
to be large, for the large deviation approximation to be good. This
is in correspondence with the results of \cite{Derrida Lebowitz},
up the fact that factor $tN^{-3/2}$ is corrected by the prefactor
$a,$ which decays as $a\sim\lambda^{-1/4}$ as $\lambda\to\infty.$
When $\lambda$ grows with $N$ as $\mbox{\ensuremath{\lambda\sim}}N^{\alpha}$,
we expect that the dynamical exponent varies continuously as $z=3/2+\alpha/4$
from KPZ, $z=3/2$, to diffusive, $z=2,$ value as $\alpha$ varies
from $\alpha=0$ to $\alpha=2$. As was discussed above, the method
of asymptotical analysis used is valid also for $\lambda$ growing
with $N$ slower than $N^{2}.$ Hence, the applicability of the Derrida-Lebowitz
scaling form of the LDF extends to systems with larger than KPZ characteristic
time scales, until the scaling becomes diffusive, $t\sim N^{2}$.
In the DA limit all particles stick together into the cluster, which
performs an ordinary random walk making $M$-step jumps at a time.
The central part of its LDF is expected to be pure Gaussian $\mbox{Prob}\left(Y_{t}/t<y\right)\sim\exp\left(-tM^{-2}y^{2}/2\right).$
This  indicates that the transition regime is expected on the scale
$\lambda\sim N^{2}.$ 

The cumulants of $Y_{t}$ can be obtained by differentiating the function
(\ref{eq:ldf-kpz}). In particular, from the first derivative we obtain
the mean number of particle jumps per unit time up to the first order
finite size correction that was already obtained in (\ref{eq:current correctioon (asymp)}).
Note that it is the value of this correction, 
\begin{equation}
\left(j^{ASEP}(L)-j^{ASEP}((\infty)\right)L=ab,\label{eq:current correction -2}
\end{equation}
which is the denominator of the argument of the LDF $\hat{G}$ in
(\ref{eq:ldf}). As noted in \cite{Derrida Lebowitz} the applicability
of the scaling form of the LDF obtained is limited by the condition
that the argument of the function $\widehat{G}(x)$ is of the order
of one. Therefore, the denominator plays the role of the scale for
the deviation of time-averaged number of jumps from its mean value,
in which the scaling form (\ref{eq:ldf}) is valid. Remarkably, its
value stays finite when $\lambda$ grows to infinity. The diffusion
coefficient for one particle, related to the second cumulant, is 
\begin{equation}
\Delta=\frac{\left(1-c\right)^{3/2}}{c^{2}}\frac{b^{2}a}{2\sqrt{2L}},\label{eq:Delta exact}
\end{equation}
decaying as $L^{-1/2}$, which is specific for the KPZ class. In the
limit $\lambda\to\infty,$ we have 
\begin{equation}
\Delta\simeq\frac{\lambda^{1/4}}{\sqrt{L}}\frac{3}{4}\sqrt{\frac{\pi}{2}}\frac{p(1-p)}{\left[c(1-c)\right]^{1/4}},\label{eq:Delta lambda->infty}
\end{equation}
which again signals that when $\lambda\sim L^{2}$ the motion of particles
changes from subdiffusive to diffusive. However, as it was discussed
above, the saddle point method fails at this scale of $\lambda$ and
we should again use different asymptotic analysis. 
\begin{figure*}[t]
\begin{tabular}{ccc}
\includegraphics[width=0.3\textwidth]{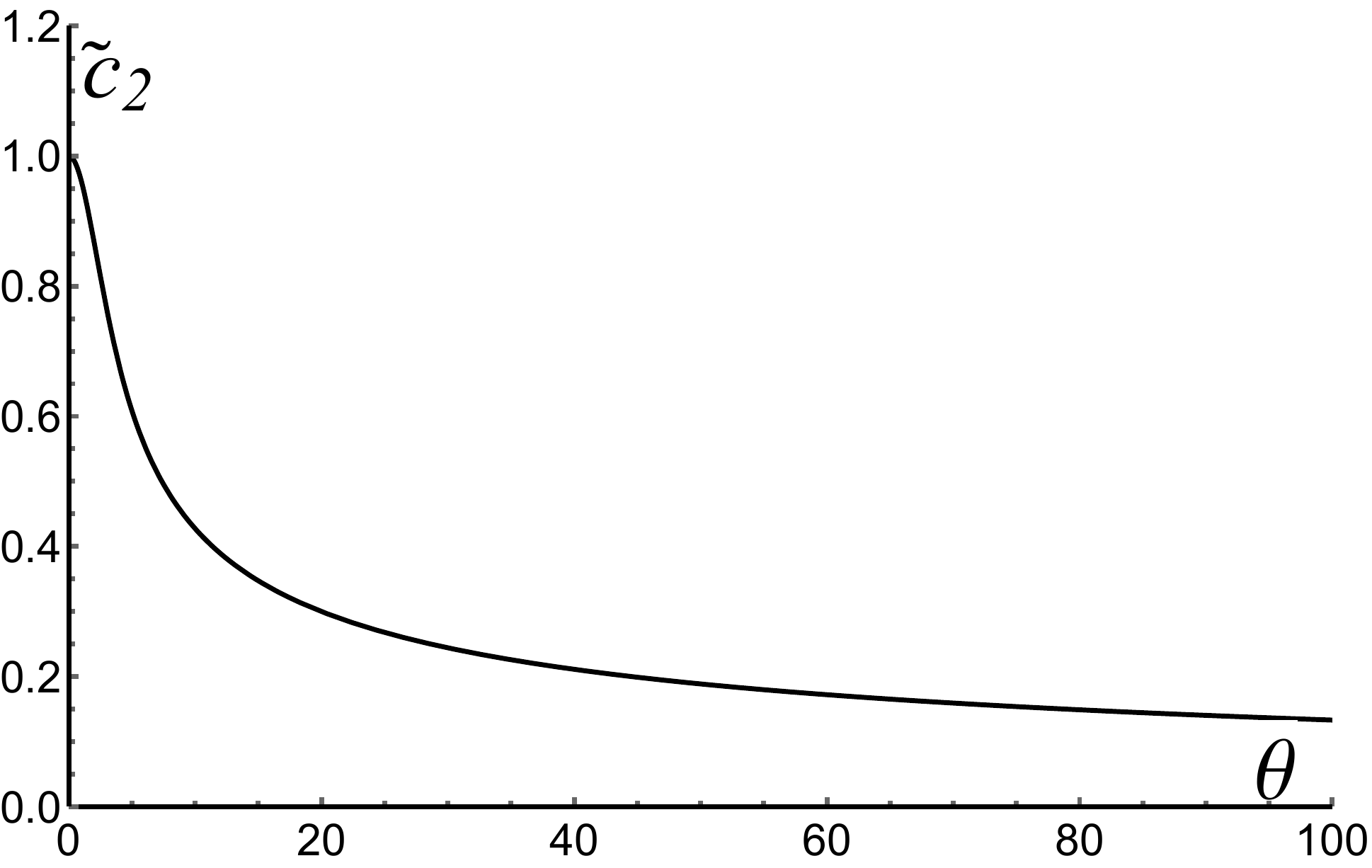}  & \includegraphics[width=0.3\textwidth]{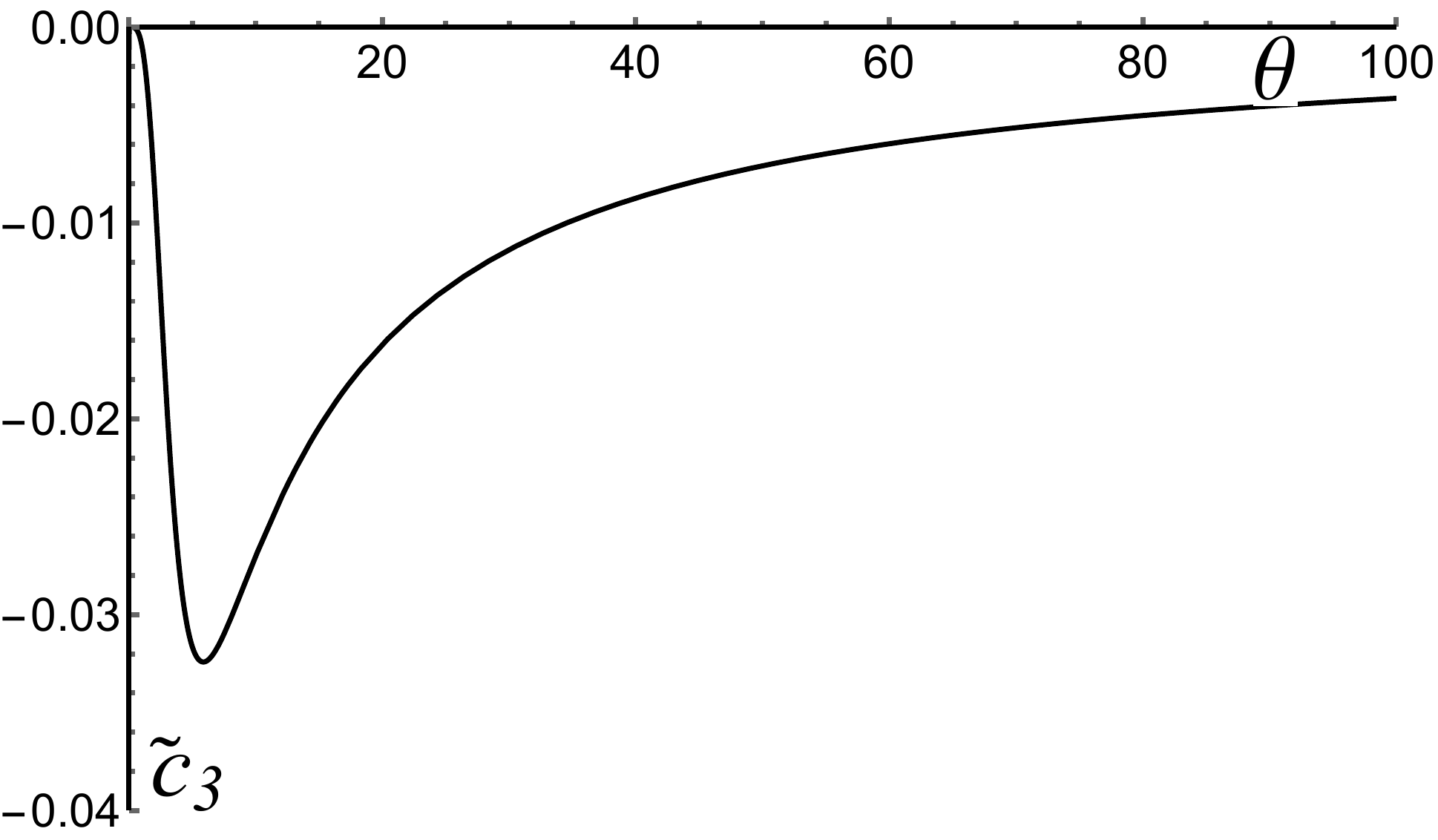} & \includegraphics[width=0.3\textwidth]{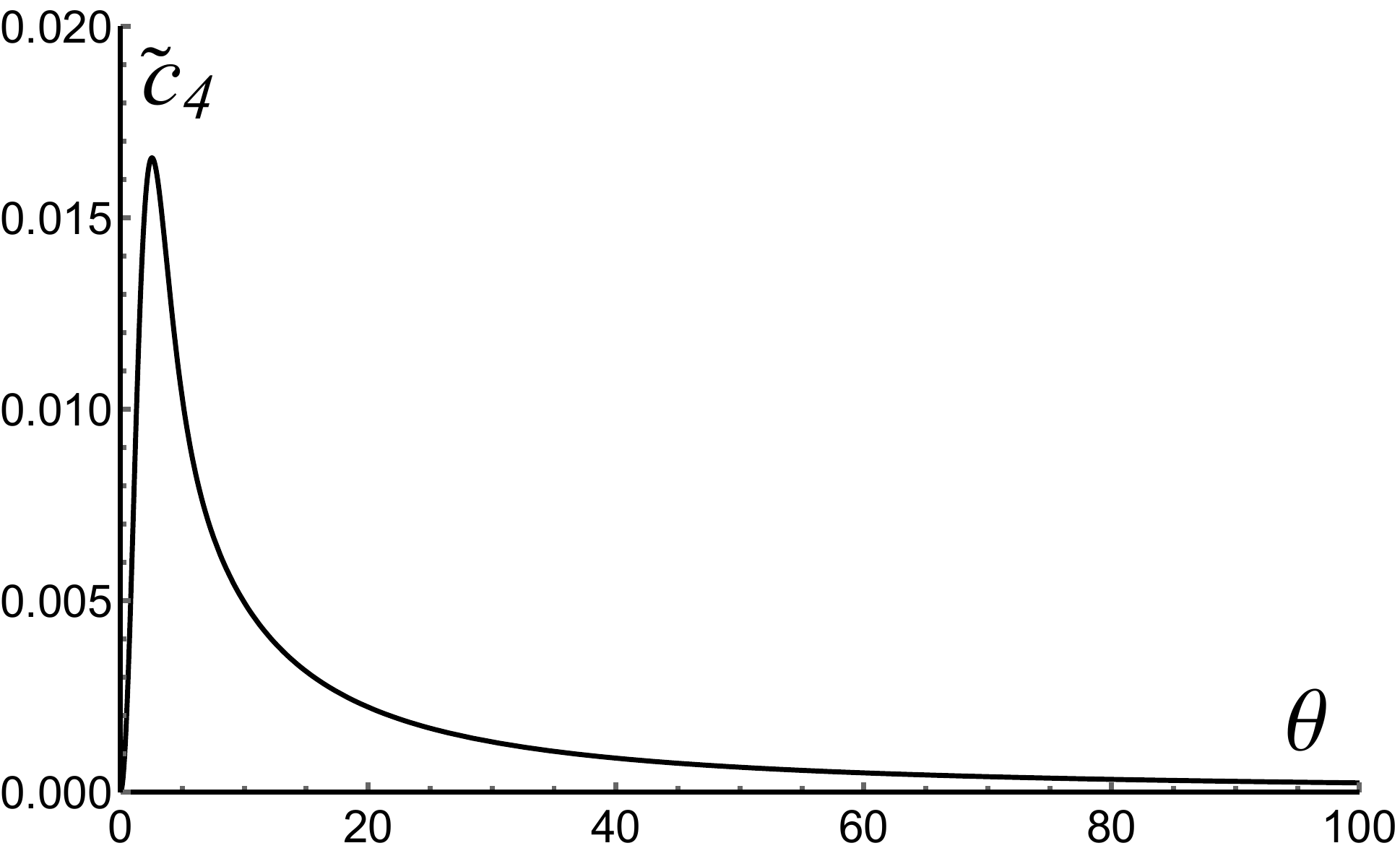}\tabularnewline
\end{tabular}

\caption{The rescaled cumulants of total number of particle jumps $\tilde{c}_{n}$
vs $\theta$ in the transition regime $N\to\infty,$ $\lambda N^{-2}=const$
.\label{fig:The-rescaled-cumulants}}
\end{figure*}

\subsubsection{The transition regime: $\lambda/N^{2}=const$, $\gamma N^{2}=const$}

When $\lambda\sim N^{2}$, the integrals $\mathcal{I}_{nN}(h(z),r(z))$
and $\mathcal{I}_{nN}(h(z),s(z))$ can be evaluated in terms of the
modified Bessel functions of the first kind as in the second part
of Sec. \ref{sub:Asymptotic-analysis}. As a result we obtain the
parametrically defined function $\ln\Lambda(\gamma)$

\begin{eqnarray}
\ln\Lambda(\gamma) & =\gamma pM+N^{-2} & p(1-p)\mathcal{G}_{\theta}(N^{2}\rho\gamma),\label{eq:Lambda - transition}
\end{eqnarray}
where the function $\mathcal{G}_{\theta}\left(t\right)$ depending
on the transition parameter $\theta$ defined in (\ref{eq:theta})
has the following parametric form: 
\[
\mathcal{G}_{\theta}\left(t\right)=\frac{\theta^{2}}{4}\sum_{k=1}^{\infty}I_{2}(k\theta)\frac{B^{k}}{k},\,\,\, t=-\frac{\theta}{2}\sum_{k=1}^{\infty}I_{1}(k\theta)\frac{B^{k}}{k}.
\]
To observe the nontrivial scaling, $\gamma$ must scale with $N$
so that the limit $\gamma N^{2}=const$ holds.

Using an asymptotic form of the modified Bessel functions (\ref{eq:Bessel large x})
for large $\theta$ and small $t$, such that $t^{2}\theta=const,$
we find 
\[
\mathcal{G_{\theta}}(t)\simeq-\frac{\theta t}{2}+\frac{3}{8}\sqrt{\frac{\theta}{2\pi}}G\left(t\sqrt{\frac{8\pi}{\theta}}\right),\,\,\,\theta\to\infty
\]
where $G(x)$ is the Derrida-Lebowitz scaling function (\ref{eq:Der-Lib}).

In the opposite DA limit the particles in a finite system form a single
cluster of $M$ particles, which move together performing the Bernoulli
random walk. In this case the exact cumulant generating function is
\begin{eqnarray*}
\lim_{t\to\infty}t^{-1}\ln\left\langle e^{\gamma Y_{t}}\right\rangle  & = & \ln\left(1-p+pe^{\gamma M}\right).\\
 & \simeq & Mp\gamma+M^{2}p(1-p)\frac{\gamma^{2}}{2},
\end{eqnarray*}
where in the second line we show the two first terms of the small
$\gamma$ expansion. These are the only terms responsible for the
limit $\lim_{M\to\infty}M^{2}\left(\lim_{t\to\infty}t^{-1}\ln\left\langle e^{\gamma Y_{t}}\right\rangle -Mp\gamma\right)=M^{4}\gamma^{2}$
to exist under condition $\gamma M^{2}=const$. This agrees with the
behavior of (\ref{eq:Lambda - transition}) at small $\theta$, which
follows from the limiting form of function $\mathcal{G}_{\theta}\left(t\right)$:
\[
\mathcal{G}_{\theta}\left(t\right)\simeq\frac{t^{2}}{2}-\frac{\theta^{2}t}{8},\,\,\,\theta\to0.
\]
The distribution of the time-averaged number of particle jumps corresponding
to (\ref{eq:Lambda - transition}) has the form 
\[
\mbox{Prob}\left(Y_{t}/t<y\right)\sim\exp\left[-\frac{p(1-p)t}{N^{2}}\mathcal{\widehat{G}}_{\theta}\left(\frac{y-Mp}{\rho p(1-p)}\right)\right],
\]
where LDF $\mathcal{\widehat{G}}_{\theta}(x)$ is the Legendre transform
of $\mathcal{G}_{\theta}\left(t\right).$ The presence of the factor
$tN^{-2}$ is specific for the diffusive systems, though the LDF has
a nontrivial form, unlike purely quadratic Gaussian single particle
case. It follows from the above analysis of the limiting behavior
of $\mathcal{G}_{\theta}\left(t\right)$ that the LDF $\mathcal{\widehat{G}}_{\theta}(x)$
continuously interpolates between the Gaussian quadratic and the KPZ
scaling function as $\theta$ varies from zero to infinity.

Differentiating $\Lambda(\gamma)$ we obtain the cumulants of this
distribution. The first one is 
\[
J\simeq Mp-p(1-p)\rho\frac{\theta}{2}\frac{I_{2}(\theta)}{I_{1}(\theta)},
\]
i.e. a finite number of macroscopic clusters present on the lattice
for a finite fraction of time results in a finite correction to the
total number of jumps, which exactly equals to $Mp$ in the DA limit.
In the small-$\theta$ limit this fraction is approximately $\theta^{2}/2$.
Making a small-$\theta$ expansion one can see that this contribution,
$p(1-p)\rho\theta^{2}/8$, is indeed of the same order also depending
on particle density.

The one-particle diffusion coefficient obtained from the second cumulant,

\begin{equation}
\Delta=p(1-p)\left[\frac{I_{1}(2\theta)}{I_{1}^{2}(\theta)}\left(\frac{I_{2}(2\theta)}{I_{1}(2\theta)}-\frac{I_{2}(\theta)}{I_{1}(\theta)}\right)\right],\label{1}
\end{equation}
is finite in the thermodynamic limit, similarly to the one-particle
random walk, when it exactly equals $p(1-p)$. In the KPZ-DA transition
regime, this value is corrected by the factor in the square brackets.
In the limit $\theta\to0$ this factor saturates to one, recovering
the free-particle diffusion coefficient. As $\theta\to\infty$ the
diffusion coefficient behaves as $\Delta\simeq(3p(1-p)/4)\sqrt{\pi/\theta}$
indicating the transition to KPZ behavior. What is different from
the DA limit as well as from the KPZ regime is the behavior of the
higher cumulants. As follows from the formula (\ref{eq:Lambda - transition})
they have the scaling $c_{n}\sim N^{2(n-1)}$ unlike $c_{n}\sim N^{3/2(n-1)}$
in the KPZ regime and $c_{n}\sim N^{n}$ in the DA limit. It is remarkable
that in the transition regime the order of growth of the cumulants
with $N$ is higher than in both the KPZ and the DA limit for $n\geq3$.
Then, it is natural to expect that the cumulants, rescaled to remove
the dependence on the system size and all the other parameters except
$\theta$, 
\[
\tilde{c}_{n}=\lim_{N\to\infty}c_{n}\times\left[N^{2(n-1)}\rho^{n}p(1-p)\right]^{-1}=\mathcal{G}_{\theta}^{(n)}(0)
\]
will vanish in both $\theta\to0$ and $\theta\to\infty$ limits having
an extrema at some finite values of $\theta$. Indeed, as seen from
Fig.~\ref{fig:The-rescaled-cumulants} the third and fourth rescaled
cumulants show the non-monotonous behavior having minimum and maximum
at some finite values of $\theta,$ respectively. The quantity, which
can be used as a measure of proximity to the KPZ regime, is the universal
cumulant ratio, 
\[
R(\theta)=\frac{c_{3}^{2}}{c_{2}c_{4}}=\frac{\left(\mathcal{G}_{\theta}^{(3)}(0)\right)^{2}}{\mathcal{G}_{\theta}^{''}(0)\mathcal{G}_{\theta}^{(4)}(0)},
\]
\begin{figure}
\includegraphics[width=0.9\columnwidth]{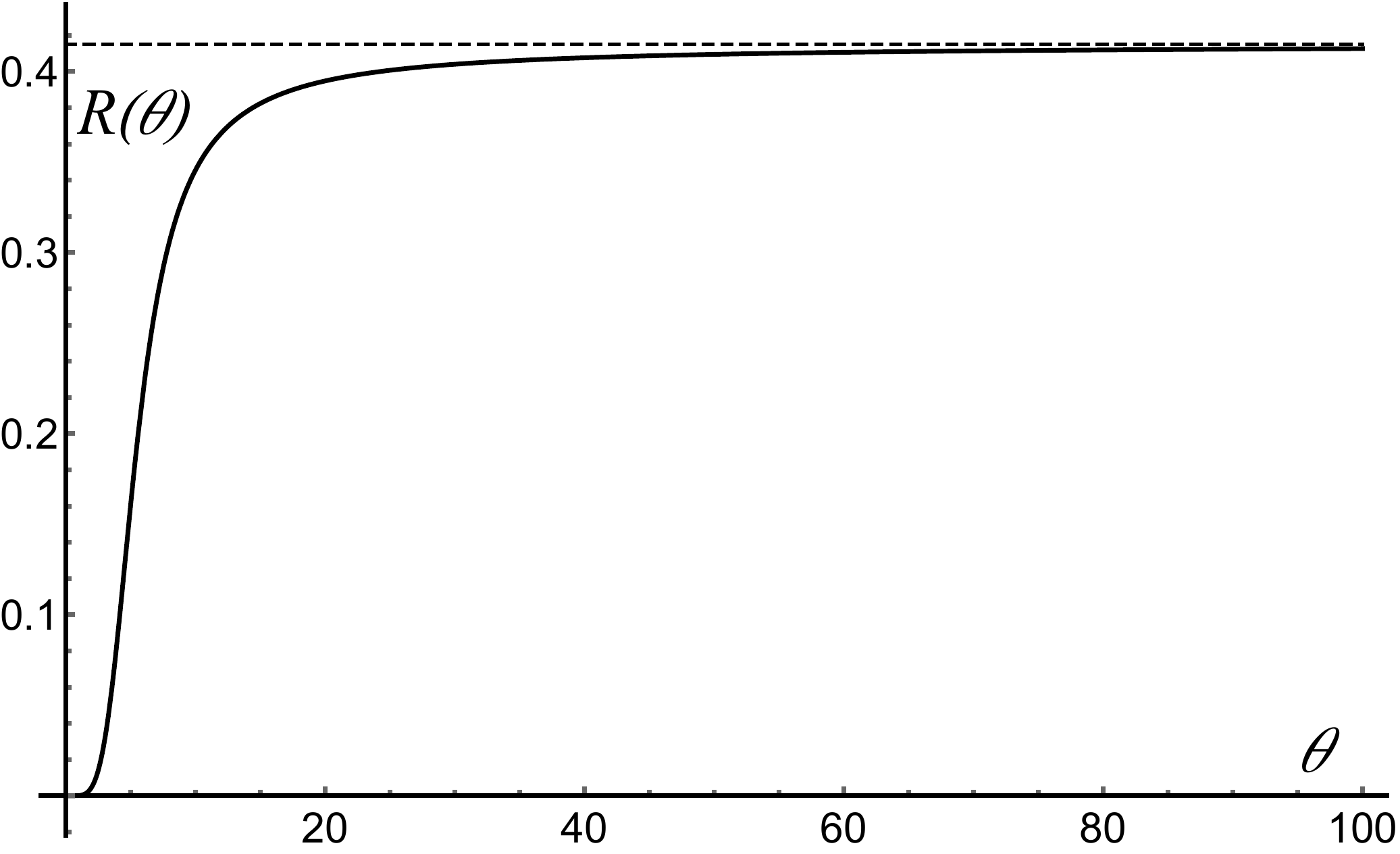}\caption{Universal cumulant ratio $R(\theta).$ The dashed line shows the limiting
KPZ value $R(\infty).$\label{fig:Universal-cumulant-ratio}}
\end{figure}
depending solely on the parameter $\theta$. As shown in Fig.~\ref{fig:Universal-cumulant-ratio},
starting from zero at $\theta=0$ the ratio $R(\theta)$ monotonously
approaches its limiting universal KPZ value 
\[
\lim_{\theta\to\infty}R(\theta)=\frac{2\left(3/2-8/3^{3/2}\right)^{2}}{15/2-24/\sqrt{3}+9/\sqrt{2}}\simeq0.41517,
\]
first obtained in \cite{Derrida Appert}.

\section{Universality and relation to KPZ equation \label{sec:Universality-and-relation}}

In context of stochastic models, the concept of universality suggests
that in the scaling limit a large class of models is characterized
by probability distributions having the same universal functional
form. The notion of the scaling limit implies that the temporal and
spacial coordinates as well as the random variables of interest are
measured in scales related to each other via simple power laws. Their
exponents, usually referred to as critical exponents, is a fixed set
of numbers specifying given universality class. In this way the scales
are defined up to non-universal constants, which depend on parameters
of the specific model. Correspondingly, numerical quantities characterizing
the random variables, e.g., cumulants or correlators, depend on these
constants only. As applied to the problem of KPZ interface growth
in one-dimensional system of size $L$, the distribution of the height
$h(x,t)$ of growing interface being a random function of the spatial
and temporal coordinates $x$ and $t$ is characterized by two sets
of amplitudes \cite{krug meakin halpin-healy},

\begin{eqnarray}
a_{n} & = & \lim_{t\to\infty}\lim_{L\to\infty}t^{-n/3}\left\langle \left(h(x,t)-\overline{h}\right)^{n}\right\rangle _{c}\label{eq:a_n}\\
b_{n} & = & \lim_{L\to\infty}\lim_{t\to\infty}L^{-n/2}\left\langle \left(h(x,t)-\overline{h}\right)^{n}\right\rangle _{c}\label{eq:b_n}
\end{eqnarray}
for transient, $t\ll L^{3/2}$, and stationary, $t\gg L^{3/2},$ parts
of evolution respectively, where $n\in\mathbb{N}$, $\left\langle x^{n}\right\rangle _{c}$
is the notation for $n$-th cumulant of the random variable $x$,
and $\bar{h}=L^{-1}\int_{0}^{L}h(x,t)dx$ is the mean interface height
for a given process realization. Also one can define finite time (size)
corrections to the average interface velocity as compared to the one
calculated at infinite time (in infinite system), 
\begin{eqnarray}
a_{v} & = & \lim_{t\to\infty}\lim_{L\to\infty}t^{2/3}\left(\left\langle \partial h/\partial t\right\rangle -v_{\infty}\right)\label{eq:a_v}\\
b_{v} & = & \lim_{L\to\infty}\lim_{t\to\infty}L\left(\left\langle \partial h/\partial t\right\rangle -v_{\infty}\right),\label{eq:b_v}
\end{eqnarray}
where $v_{\infty}=\lim_{t,L\to\infty}\left\langle \partial h/\partial t\right\rangle .$
It was conjectured in \cite{krug meakin halpin-healy} that all these
quantities can be expressed in terms of only two dimensional invariants.
The conjecture was first proposed based on analysis of the KPZ equation
itself, 
\begin{equation}
\frac{\partial h}{\partial t}=\widetilde{\nu}\Delta h+\widetilde{\lambda}\left(\nabla h\right)^{2}+\eta,\label{eq:KPZ}
\end{equation}
which was the first prototypical model catching the universal features
of the KPZ class. Here the notations for parameters $\widetilde{\lambda}$
and $\widetilde{\nu}$ have a tilde to keep the notations traditional
for KPZ equation and distinguish them from the $\nu$ and $\lambda$
of our model. The white noise $\eta$ is fully characterized by the
covariance 
\[
\left\langle \eta(x,t)\eta(x',t')\right\rangle =D\delta(x-x')\delta(t-t').
\]
For an interface described by (\ref{eq:KPZ}) the two mentioned dimensional
invariants are $\widetilde{\lambda}$ and $A=D/2\widetilde{\nu}.$
In terms of these constants the transient amplitudes (\ref{eq:a_n})
and (\ref{eq:a_v}) are given by 
\[
a_{n}=\left(\left|\widetilde{\lambda}\right|A^{2}\right)^{n/3}\widetilde{a}_{n}\,\,\,\mbox{and}\,\,\, a_{v}=(|\widetilde{\lambda}|A^{2})\widetilde{a}_{v},
\]
where $\widetilde{a}_{n}$ and $\widetilde{a}_{v}$ are universal
numbers. These numbers, as known from the later development of the
field, must be related to cumulants of the universal distributions
like the Tracy-Widom distributions, dependent on global form (large-scale)
of initial conditions. For the stationary amplitudes (\ref{eq:b_n})
and (\ref{eq:b_v}) in the system with periodic boundary conditions
we have 
\begin{equation}
b_{2}=\frac{A}{12},\,\,\, b_{n}=0,\,\,\, n>2\label{eq:b_2}
\end{equation}
and 
\begin{equation}
b_{v}=-\frac{A\widetilde{\lambda}}{2}.\label{eq:b_v univ}
\end{equation}
Vanishing of all amplitudes $b_{n}$ except the second one is due
to the Gaussian stationary height distribution of the KPZ interface.
The universality conjectured in~\cite{krug meakin halpin-healy}
suggests that for an interface belonging to the KPZ class, the amplitudes
(\ref{eq:a_n})--(\ref{eq:b_v}) have the same dependence on $\widetilde{\lambda}$
and $A$, which can in general be defined without appealing to the
KPZ equation and measured experimentally. Namely, the parameter $\widetilde{\lambda}$,
related to the response of the interface velocity to introducing a
small tilt $h(x,t)\to h(x,t)+\kappa x$, is defined as 
\begin{equation}
\widetilde{\mbox{\ensuremath{\lambda}}}=\frac{\partial^{2}v_{\infty}}{\partial\kappa^{2}},\label{eq:lambda-def}
\end{equation}
and the parameter $A$ is the amplitude of spacial correlation function
\begin{equation}
\lim_{t\to\infty}\left\langle \left(h(x,t)-h(y,t)\right)^{2}\right\rangle _{c}=A\left|x-y\right|.\label{eq:A-def}
\end{equation}

The above studied ASEP-like system can be related to the KPZ interface
on the lattice by 
\[
h_{i+1}-h_{i}=1-2\tau_{i},
\]
where $\tau_{i}=0,1$ is the occupation number of the $i$th site
and $h_{i}$ is the interface height above the bond connecting sites
$i-1$ and $i$ of the lattice, $i=1,\dots,L.$ For this mapping being
consistent with the number of particles on the lattice the interface
must satisfy helicoidal boundary conditions 
\[
h_{i+L}=h_{i}-(L-2M),
\]
which gives a tilt $\kappa=1-2c$ to the interface. Then the change
of the interface height $\left(h_{i}(t)-h_{i}(0)\right)$ in time
is nothing but twice the number of particles that have traversed the
bond $(i-1,i)$ by the time $t$. Correspondingly the limiting interface
speed is twice the particle current in the ASEP-like system, $v_{\infty}=2j^{ASEP}$,
where $j^{ASEP}$ was obtained in (\ref{eq:j^TASEP}). Then we have
\[
\widetilde{\lambda}=\frac{1}{2}\frac{\partial^{2}j^{ASEP}}{\partial c^{2}}.
\]
Using (\ref{eq:covariance}) we obtain for $i\ll j$ 
\begin{eqnarray*}
\left\langle \left(h_{i}-h_{j}\right)^{2}\right\rangle _{c} & = & 4\sum_{i\leq k,l\leq j+1}\left(\left\langle \tau_{k}\tau_{l}\right\rangle -c^{2}\right)\\
 & \simeq & 4c(1-c)\coth\left(\frac{1}{2\xi}\right)\left|i-j\right|;
\end{eqnarray*}
i.e., $A=4c(1-c)\coth(1/(2\xi))$, where $\xi$ is the correlation
length (\ref{eq:corr length}). The explicit expressions of $\widetilde{\lambda}$
and $A$ can be found in Appendix \ref{sec:Explicit-expressions-of}.
Also the finite size correction $b_{v}$ to the limiting interface
velocity is twice the correction to the particle current given in
(\ref{eq:J asymp}) and (\ref{eq:current correction -2}). One confirmation
of the universality is the observation that the relation (\ref{eq:b_v})
between $b_{v}$ and the parameters $\widetilde{\lambda}$ and $A$,
defined by (\ref{eq:lambda-def}) and (\ref{eq:A-def}) respectively,
holds exactly (see appendix \ref{sec:Explicit-expressions-of}.).

Another demonstration of universality can be obtained using the results
of Sec. \ref{sec:Statistics-of-particle}. Note that the amplitudes
(\ref{eq:a_n})--(\ref{eq:b_v}) characterize the form of the interface
relative to its average position $\bar{h}.$ At the same time the
absolute value of the interface height is dominated by the position
of its center of mass, which, up to the bounded initial value, is
$\overline{h}\simeq2L^{-1}Y_{t}$. Therefore, the universal LDF obtained
for $Y_{t}$ also characterizes the statistics of the motion of the
center of mass of interface. On the other hand, its scaling properties
are expected to be defined by the dimensionful invariants $\widetilde{\lambda}$
and $A$ solely. In particular, simple dimensional arguments together
with the scaling ansatz show~\cite{Krug review} that the variance
of $\overline{h},$ related to the diffusion coefficient of a particle
by $\left\langle \overline{h}^{2}\right\rangle _{c}=4c^{2}\Delta t,$
has the form 
\begin{equation}
\left\langle \overline{h}^{2}\right\rangle _{c}=s_{0}A^{3/2}|\widetilde{\lambda}|L^{-1/2}t,\label{eq:h bar universal}
\end{equation}
where $s_{0}=\sqrt{\pi}/4$ is a universal number first obtained in~\cite{DEM}.
Comparing this formula with the expression (\ref{eq:Delta exact})
obtained for $\Delta$, we obtain relation (\ref{eq:b(A,c)}) between
$b$, $A$, and the density $c$, which is indeed confirmed from explicit
calculations in Appendix \ref{sec:Explicit-expressions-of}. The same
arguments can be applied to the cumulants of an arbitrary order. In
general, all the model dependence of the scaled cumulant generating
function (and hence of the LDF) obtained in Sec. \ref{sec:Statistics-of-particle}
is incorporated into two constants $a$ and $b$. It takes some algebra
to show that these constants can be reexpressed in terms of the dimensional
invariants $A$ and $\widetilde{\lambda}$ of this section: 
\begin{eqnarray}
a & = & \frac{\sqrt{2A}\left|\widetilde{\lambda}\right|(1-c)^{3/2}}{4\sqrt{\pi}},\label{eq:a(A,lambda)}\\
b & \mbox{=}- & \mbox{sgn}\,\widetilde{\lambda}\,\frac{\sqrt{\pi A/2}}{(1-c)^{3/2}}.\label{eq:b(A,lambda)}
\end{eqnarray}
In the spirit of universality we conjecture this relation to be universal.
Up to our knowledge, it did not yet explicitly appear in the literature.

Finally it is informative to see how the system approaches the DA
limit. As $\lambda\to\infty$, we asymptotically have 
\[
A\simeq8\sqrt{\lambda}\left[c(1-c)\right]^{3/2}\,\,\,\mbox{and\,\,\,}\widetilde{\lambda}\simeq-\frac{3(1-p)p}{8(1-c)^{5/2}c^{1/2}}\sqrt{\frac{1}{\lambda}}.
\]
As we saw, the KPZ regime (in particular, the universal scaling form
of the LDF of interface height) holds until the value of $\lambda$
becomes of the order of $\lambda\sim N^{2},$ i.e., $A$ and $\widetilde{\lambda}$
being of order $L$ and $1/L$, respectively. Remarkably the product
$\widetilde{\lambda}A$ proportional to $b_{v}$, which is related
to the typical fluctuation range, stays finite in the limit $\lambda\to\infty$.
Therefore, first, up to the scale $\lambda\sim N^{2},$ the increase
of $\lambda$ affects only non-universal constants preserving the
universal functional form (\ref{eq:ldf-kpz}) of the LDF. Then, in
the scale $\lambda\sim N^{2}$ the functional form of LDF starts to
gradually change until reaching the purely Gaussian form. 
\begin{acknowledgments}
We acknowledge the financial support from the Government of the Russian
Federation within the framework of the implementation of the 5-100
Programme Roadmap of the National Research University Higher School
of Economics. The work was also supported by the RFBR grant under
Project 14-01-00474-a and by Heisenberg-Landau program.
\end{acknowledgments}
\appendix

\section{Explicit expressions of model-dependent constants and universal relations\label{sec:Explicit-expressions-of}}

Though the scaling functions found from the asymptotic analysis are
of a rather simple form, the model-dependent constants expressed as
functions of particle density are very cumbersome. It is much more
efficient to consider them as functions of associated fugacities.
In particular, the fugacity $z_{-}$ appears in the analysis of the
stationary measure of the ZRP-like system in Sec. \ref{sub:Asymptotic-analysis}
as the saddle point, where the main contribution to the integrals
comes from. Its relation to the particle density in the TASEP-like
system is given in (\ref{eq:z_pm}), or conversely 
\begin{equation}
c=\frac{(1-\nu)z_{-}}{1-\nu\left(2-z_{-}\right)z_{-}}.\label{eq:c(z_)}
\end{equation}
Then we obtain the mean number of particle jumps per unit time, 

\begin{align}
J & =N\frac{z_{-}(\mu-\nu)}{\left(1-\mu z_{-}\right)\left(1-\nu z_{-}\right)}\\
 & +\frac{(1-\mu)(\mu-\nu)}{(1-\nu)}\frac{\left(1-z_{-}\right)z_{-}\left(1-\nu z_{-}\right)\left(1-\mu\nu z_{-}^{3}\right)}{\left(1-\mu z_{-}\right){}^{3}\left(1-\nu z_{-}^{2}\right){}^{2}},\nonumber 
\end{align}
which being divided by size of the system $L$, yields the thermodynamic
value of the mean particle current, 
\begin{align}
j_{\infty} & =(1-c)\frac{z_{-}(\mu-\nu)}{\left(1-\mu z_{-}\right)\left(1-\nu z_{-}\right)}\\
 & =\frac{(\mu-\nu)\left(1-z_{-}\right)z_{-}}{\left(1-\mu z_{-}\right)\left(1-\nu\left(2-z_{-}\right)z_{-}\right)},\nonumber 
\end{align}
plus the $1/L$ finite-size correction

\begin{align}
L(j_{L}-j_{\infty}) & =\frac{(1-\mu)(\mu-\nu)}{(1-\nu)}\label{eq: current correction (z_)}\\
\times & \frac{\left(1-z_{-}\right)z_{-}\left(1-\nu z_{-}\right)\left(1-\mu\nu z_{-}^{3}\right)}{\left(1-\mu z_{-}\right){}^{3}\left(1-\nu z_{-}^{2}\right){}^{2}}.\nonumber 
\end{align}
These expressions show up again in Sec. \ref{sub:Scaling-limits},
where the latter one appears to be a product of the two model-dependent
constants,

\begin{align}
a & =\frac{(1-\mu)(\mu-\nu)}{\sqrt{2\pi}(1-\nu)^{3/2}}\label{eq:a(z_)}\\
 & \times\frac{\sqrt{z_{-}}\left(1-z_{-}\right){}^{2}\left(1-\nu z_{-}\right){}^{2}\left(1-\mu\nu z_{-}^{3}\right)}{\left(1-\mu z_{-}\right){}^{3}\left(1-\nu z_{-}^{2}\right){}^{5/2}},\nonumber \\
\nonumber \\
b & =\frac{\sqrt{2\pi(1-\nu)z_{-}\left(1-\nu z_{-}^{2}\right)}}{\left(1-z_{-}\right)\left(1-\nu z_{-}\right)},\label{eq:b(z_)}
\end{align}
which also determine the scaling behavior of all higher cumulants,
such as diffusion coefficient, as well as the characteristic temporal
and fluctuation scales. Indeed, multiplying (\ref{eq:a(z_)}) and
(\ref{eq:b(z_)}) we obtain exactly (\ref{eq: current correction (z_)})
confirming the announced relation (\ref{eq:current correction -2}). 

Another fugacity $z^{*},$ given in (\ref{eq:z^*}), appears in Sec.
\ref{sub:Transfer-matrix-approach} from transfer-matrix analysis
of the TASEP-like system. The two seemingly unrelated sets of results
obtained from ZRP-like and TASEP-like systems turn out to be linked,
when one checks the universal relations between the two-dimensional
invariants obtained in KPZ theory. One of the invariants, the amplitude
of the correlation function (\ref{eq:A-def}), is given in terms of
$z^{*}$ by 
\begin{align}
A & =4c(1-c)\frac{\lambda_{1}+\lambda_{2}}{\lambda_{1}-\lambda_{2}}\label{eq:A}\\
 & =\frac{4(1-\nu)z^{*}(z^{*}+1)}{\left((z^{*}+1)^{2}-4\nu z^{*}\right)^{3/2}},\nonumber 
\end{align}
while the other one, the non-linearity coefficient from KPZ equation,
obtained from the second derivative of particle current depending
on $z_{-}$:

\begin{align}
\widetilde{\lambda} & =\frac{1}{2}\frac{\partial^{2}j_{\infty}}{\partial c^{2}}=\frac{1}{2}\left(\frac{1}{\partial c/\partial z_{-}}\frac{\partial}{\partial z_{-}}\right)^{2}j_{\infty}\\
 & =-\frac{(1-\mu)(\mu-\nu)}{(1-\nu)^{2}}\frac{\left(1-\nu\left(2-z_{-}\right)z_{-}\right){}^{3}\left(1-\mu\nu z_{-}^{3}\right)}{\left(1-\mu z_{-}\right){}^{3}\left(1-\nu z_{-}^{2}\right){}^{3}}.\nonumber 
\end{align}
Noting that the fugacities are related by 
\begin{equation}
z^{*}=\frac{z_{-}\left(\nu z_{-}-1\right)}{z_{-}-1}\label{eq:z^* vs z_}
\end{equation}
we find that the product of the dimensional invariants is equal to
\[
\widetilde{\lambda}A=-2b_{v},
\]
where $b_{v}=2ab=2L(j_{L}-j_{\infty}),$ which is nothing but the
relation (\ref{eq:b_v}). Another relation, 

\begin{equation}
b=\frac{\sqrt{\pi A}}{\sqrt{2}\left(1-c\right){}^{3/2}},\label{eq:b(A,c)}
\end{equation}
can be verified by direct examining formulas (\ref{eq:c(z_)},\ref{eq:b(z_)},\ref{eq:A},\ref{eq:z^* vs z_}).
This complies with another prediction of KPZ theory (\ref{eq:h bar universal})
in conjunction with the formula (\ref{eq:Delta exact}) obtained for
the diffusion coefficient. Then the connection (\ref{eq:a(A,lambda)},\ref{eq:b(A,lambda)})
between $a,b$ and $\widetilde{\lambda},A$ is straightforward.

\section{Hypergeometric functions\label{sec:Hypergeometric-functions} }

\subsection{Gauss hypergeometric functions}

Series representation: 
\[
_{2}F_{1}(a,b;c;x)=\sum_{n=0}^{\infty}\frac{(a)_{n}(b)_{n}}{(c)_{n}n!}x^{n}
\]
Generating function for terminating series $_{2}F_{1}(a,b;c;x)$ with
$a$ negative integer: 
\begin{eqnarray*}
G(x,t) & \equiv & \frac{(1-xt)^{\alpha}}{(1-t)^{\beta}}\\
 & = & \sum_{n=0}^{\infty}\frac{(\beta)_{n}}{n!}\left._{2}F_{1}\right.(-n,-\alpha,-\beta-n+1;x)t^{n}
\end{eqnarray*}
Euler transformation: 
\begin{equation}
F(a,b;c;z)=\left(1-z\right)^{c-a-b}F(c-a,c-b;c;z)\label{eq:Euler}
\end{equation}
Chu-Vandermonde identity: 
\begin{equation}
\left._{2}F_{1}\right.(-n,-\alpha,-\beta-n+1;1)=\frac{(\beta-\alpha)_{n}}{(\beta)_{n}}\label{eq:Chu}
\end{equation}

\subsection{Appell hypergeometric function $F_{1}$}

Series representations: 
\begin{eqnarray*}
F_{1}(\alpha;\beta,\beta';\gamma;x,y) & = & \sum_{n,m=0}^{\infty}\frac{(\alpha)_{m+n}(\beta)_{m}(\beta')_{n}}{(\gamma)_{m+n}m!n!}x^{m}y^{n}
\end{eqnarray*}
Generating function for terminating series $F_{1}(\alpha;\beta,\beta';\gamma;x,y)$
with $\alpha$ negative integer : 
\begin{eqnarray*}
G(x,y,z) & \equiv & \left(1-z\right)^{\alpha}(1-xz)^{-\beta}(1-yz)^{-\beta'}\\
 & = & \sum_{n=0}^{\infty}\frac{(-\alpha)_{n}}{n!}F_{1}(-n;\beta,\beta';\alpha-n+1;x,y)z^{n}
\end{eqnarray*}

One-variable reduction:

\begin{equation}
F_{1}(\alpha;\beta,\beta';\gamma;x,0)=\left._{2}F_{1}\right.(\alpha,\beta;\gamma;x)\label{eq:F_1(0)}
\end{equation}

\begin{widetext}Generalized Chu-Vandermonde identity 
\begin{equation}
F_{1}(-n;\beta,\beta';\alpha-n+1;x,1)=\frac{\left(\beta'-\alpha\right)_{n}}{(-\alpha)_{n}}\left._{2}F_{1}\right.(-n,\beta;\alpha-\beta'-n+1;x)\label{eq:Chu-1}
\end{equation}

\end{widetext}

\section{Modified Bessel function\label{sec:Modified-Bessel-function}}

Integral representation: 
\[
I_{k}(y)=\int_{0}^{2\pi}\exp\left(y\cos\varphi+ki\varphi\right)\frac{d\varphi}{2\pi}
\]
Asymptotic behavior: 
\begin{eqnarray}
I_{\alpha}(x) & = & \frac{e^{x}}{\sqrt{2\pi x}}\Big(1-\frac{4\alpha^{2}-1}{8x}\Big),\,\,\, x\to\infty\label{eq:Bessel large x}\\
I_{\alpha}(x) & = & \frac{1}{\alpha!}\Big(\frac{x}{2}\Big)^{\alpha},\,\,\, x\to0\label{eq:Bessel small x}
\end{eqnarray}


\begin{thebibliography}{10}
\bibitem{Spoh91} H. Spohn, \textit{Large Scale Dynamics of Interacting
Particles}, (Springer, Berlin, 1991).

\bibitem{Ligg99} T.L. Liggett,\textit{ Stochastic interacting systems:
contact, voter and exclusion processes, }(Springer, Berlin, 1999)

\bibitem{Gunter} G.M. Schütz, Solvable models for many-body systems
far from equilibrium. In C.Domb and J.Lebowitz (eds.) \textit{Phase
Transitions and Critical Phenomena}, Vol.19 (Academic, London, pp.1-251,
2001).

\bibitem{EW} S.F. Edwards and D.R. Wilkinson, Proc. R. Soc. A \textbf{381},17
(1982)

\bibitem{KPZ} M. Kardar, G. Parisi, and Y. C. Zhang, Phys. Rev. Lett.
\textbf{56}, 889 (1986)

\bibitem{Gwa Spohn}L. H. Gwa and H. Spohn Phys. Rev. A \textbf{46},
844 (1992)

\bibitem{Doochul Kim} D. Kim, Phys. Rev. E \textbf{52}, 3512 (1995)

\bibitem{Derrida Lebowitz}B. Derrida and J.L. Lebowitz, Phys. Rev.
Lett. \textbf{80}, 209 (1998)

\bibitem{de Gier Essler}J. de Gier and F. H. L. Essler , Phys. Rev.
Lett. \textbf{ 107}, 010602. (2011)

\bibitem{GLMV} M. Gorissen, A. Lazarescu, K. Mallick, and C. Vanderzande,
Phys. Rev. Lett. \textbf{109}, 170601 (2012)

\bibitem{Schutz} G.M. Schütz, J. Stat. Phys. \textbf{88}, 427 (1997)

\bibitem{priezzhev} V. B. Priezzhev, Phys. Rev. Lett. \textbf{91},
050601 (2003)

\bibitem{Sasamoto} T. Sasamoto, J. Phys. A \textbf{38,} L549 (2005)

\bibitem{BFPS} A. Borodin, P.L. Ferrari, M. Pr$\ddot{\mbox{a}}$hofer,
and T. Sasamoto, J. Stat. Phys. \textbf{129} 1055 (2007)

\bibitem{IS}T. Imamura and T. Sasamoto, J. Stat. Phys. \textbf{128}
799-846 (2007)

\bibitem{BFPS1}A. Borodin, P.L. Ferrari, M. Pr$\ddot{\mbox{a}}$hofer,
and T. Sasamoto, Int. Math. Res. Papers, rpm002 (2007)

\bibitem{BF} A. Borodin and P.L. Ferrari, Electron. J. Probab. \textbf{13},
1380 (2008)

\bibitem{BFS} A. Borodin, P.L. Ferrari and T. Sasamoto, Comm. Pure
Appl. Math. \textbf{61}, 1603 (2008)

\bibitem{PovPriS}A. M. Povolotsky, V. B. Priezzhev and G. M. Schütz,
J. Stat. Phys. \textbf{142}, 754 (2011)

\bibitem{PPP}S. S. Poghosyan, A. M. Povolotsky, and V. B. Priezzhev,
J. Stat. Mech. \textbf{08} P08013 (2012)

\bibitem{BrPrS}J.G. Brankov, V.B. Priezzhev and R.V. Shelest, Phys.
Rev. E \textbf{69} 066136 (2004)

\bibitem{PP}A.M. Povolotsky and V.B. Priezzhev, J. Stat. Mech. P07002
(2006)

\bibitem{PPS} S. S. Poghosyan, V. B. Priezzhev, and G. M. Schütz,
J. Stat. Mech. P04022 (2010)

\bibitem{Woelki} M. Woelki, Steady States of discrete mass transport
models, master thesis, University of Duisburg-Essen (2005)

\bibitem{genTASEP}A.E. Derbyshev, S.S. Poghosyan, A.M. Povolotsky,
and V.B. Priezzhev, J. Stat. Mech. P05014 (2012)

\bibitem{chipping} A.M. Povolotsky, J. Phys. A \textbf{46}, 465205
(2013)

\bibitem{Corwin2}I. Corwin, Int. Math. Res. Notices rnu094 (2014)

\bibitem{BCPS}A. Borodin, I. Corwin, L. Petrov, and T. Sasamoto,
arXiv:1407.8534

\bibitem{Povolotsky Mendes} A.M. Povolotsky and J. F. F. Mendes,
J. Stat. Phys. \textbf{123}, 125 (2006)

\bibitem{lee kim} D. S. Lee and D. Kim, Phys. Rev. E \textbf{59},
6476 (1999)

\bibitem{PPH} A. M. Povolotsky, V. B. Priezzhev, and C. K. Hu, J.
Stat. Phys.\textbf{ 111}, 1149 (2003)

\bibitem{povolotsky}A. M. Povolotsky, Phys. Rev. E \textbf{69}, 061109
(2004)

\bibitem{AntalSchutz} T. Antal and G.M. Sch$\mathrm{\ddot{u}}$tz,
Phys. Rev. E \textbf{62}, 83 (2000)

\bibitem{Majumdar} S.N. Majumdar, S.Krishnamurthy, and M. Barma,
Phys. Rev. Lett. \textbf{81}, 3691 (1998)

\bibitem{krug meakin halpin-healy} J. Krug, P. Meakin, and T. Halpin-Healy,
Phys. Rev. A \textbf{45}, 638 (1992)

\bibitem{Evans Zia Majumdar} M.R. Evans, S. N. Majumdar, and R.K.P.
Zia, J. Phys. A \textbf{37}, L275 (2004)

\bibitem{evans}M. R. Evans, Braz. J. Phys. \textbf{30} 42 (2000)

\bibitem{burda}P. Bialas, Z. Burda, and D. Johnston, Nucl. Phys.
B \textbf{493} 505 (1997)

\bibitem{Brankov Papoyan Poghosyan  Priezzhev} J. G. Brankov, V.V.
Papoyan, V. S., Poghosyan, and V.B. Priezzhev, Physica A \textbf{368},
471 (2006)

\bibitem{Kanai} M. Kanai, J. Phys. A \textbf{40} 7127 (2007)

\bibitem{SSNI} M. Schreckenberg, A. Schadschneider, K. Nagel, and
N. Ito, Phys. Rev. E \textbf{51}, 2939 (1995)

\bibitem{RSSS} N. Rajewsky, L. Santen, A. Schadschneider and M. Schreckenberg
J. Stat. Phys. \textbf{92}, 151 (1998)

\bibitem{Derrida Appert} B. Derrida, C. Appert, J. Stat. Phys. \textbf{94,}
1 (1999)

\bibitem{Krug review} J. Krug, Adv. in Phys. \textbf{46}, 139 (1997)

\bibitem{DEM}B. Derrida, M.R. Evans, and D. Mukamel, J. Phys. A \textbf{26},
4911 (1993)\end{thebibliography}
\end{document}